\documentclass[twocolumn,eqsecnum,prb,aps]{revtex4-1}
\usepackage[usenames,dvipsnames]{color}
\usepackage[english]{babel}
\usepackage{epsfig}
\usepackage{graphicx,subfigure}
\usepackage{tikz}
\usetikzlibrary{calc}
\usetikzlibrary{decorations.markings}
\usetikzlibrary{decorations.pathmorphing}
\usetikzlibrary{positioning,shapes,arrows,decorations.markings,calc}
\usepackage{longtable}
\usepackage{multirow}
\usepackage{arydshln}
\usepackage{hyperref}
\usepackage{amsmath,bbm,bm,amssymb}
\usepackage[mathcal,mathscr]{eucal}
\DeclareFontFamily{OT1}{pzc}{}
\DeclareFontShape{OT1}{pzc}{m}{it}{<-> s * [1.10] pzcmi7t}{}
\DeclareMathAlphabet{\mathpzc}{OT1}{pzc}{m}{it}
\usepackage[latin1]{inputenc}
\usepackage[T1]{fontenc}
\usepackage{times}
\usepackage[Euler]{upgreek}
\usepackage{enumerate}

\usepackage{bookmark}

\usepackage{textcomp}

\begin{document}

\title{Atomic structure of metal-halide perovskites from first principles: The chicken-and-egg paradox of the organic-inorganic interaction}

\author{Jingrui Li}
\email{jingrui.li@aalto.fi}
\affiliation{Centre of Excellence in Computational Nanoscience (COMP) and Department of Applied Physics, Aalto University, P.O.Box 11100, FI-00076 AALTO, Finland}
\author{Patrick Rinke}
\affiliation{Centre of Excellence in Computational Nanoscience (COMP) and Department of Applied Physics, Aalto University, P.O.Box 11100, FI-00076 AALTO, Finland}

\begin{abstract}
  We have studied the prototype hybrid organic-inorganic perovskite $\text{CH}_3^{}\text{NH}_3^{}\text{PbI}_3^{}$ and its three close relatives, $\text{CH}_3^{}\text{NH}_3^{}\text{SnI}_3^{}$, $\text{CH}_3^{}\text{NH}_3^{}\text{PbCl}_3^{}$ and $\text{CsPbI}_3^{}$, using relativistic density function theory. The long-range van der Waals (vdW) interactions were incorporated into the Perdew-Burke-Ernzerhof (PBE) exchange-correlation functional using the Tkatchenko-Scheffler pairwise scheme. Our results reveal that hydrogen bonding, which is well described by the PBE functional, plays a decisive role for the structural parameters of these systems, including the position and orientation of the organic cation as well as the deformation of the inorganic framework. The magnitude of the inorganic-framework deformation depends sensitively on the orientation of the organic cation, and directly influences the stability of the hybrid perovskites. Our results suggest that the organic and the inorganic components complement each other: The low symmetry of the organic cation is the origin of the inorganic-framework deformation, which then aids the overall stabilization of the hybrid perovskite structure. This stabilization is indirectly affected by vdW interactions, which lead to smaller unit-cell volumes than in PBE and therefore modulate the interaction between the organic cation and the inorganic framework. The vdW-induced lattice-constant corrections are system dependent and lead to PBE+vdW lattice constants in good agreement with experiment. Further insight is gained by analysing the vdW contributions. In all iodide-based hybrid perovskites the interaction between the organic cation and the iodide anions provides the largest lattice-constant change, followed by iodine-iodine and the organic cation -- heavy-metal cation interaction. These corrections follow an almost linear dependence on the lattice constant within the range considered in our study, and are therefore approximately additive.
\end{abstract}

\maketitle
\thispagestyle{empty}

\clearpage

\section{Introduction}

Hybrid perovskite photovoltaics (HPPV) \cite{Snaith13,Green14} has surprised the photovoltaic community with its record increase in power-conversion efficiency (PCE) during the last three years \cite{NRELchart}. The photovoltaic utilization of hybrid organometal-halide perovskites, especially the prototype compound methylammonium ($\text{MA}^+\equiv\text{CH}_3^{}\text{NH}_3^+$) lead triiodide ($\text{CH}_3^{}\text{NH}_3^{}\text{PbI}_3^{}$, shortened as $\text{MAPbI}_3^{}$ hereafter), was pioneered by Kojima \textit{et al.}, who used these materials as sensitizers in dye-sensitized solar cells and obtained $3.8\%$ PCE \cite{Kojima09}. The current state-of-the-art HPPV architecture was proposed in 2012 by replacing the liquid electrolyte with solid-state hole-transporting materials, achieving $\sim\!\!10\%$ PCE \cite{KimHS12,LeeMM12}. This triggered a rapid PCE improvement, as reflected by the current PCE record of HPPVs that broke the $20\%$ mark \cite{Jeon15}. This PCE is already quite close to the best performing inorganic-based single-junction thin-film cells, such as $\text{CdTe}$. As photoactive materials, hybrid perovskites show several essential advantages in photovoltaic applications, such as band gaps close to the optimal value for single-junction solar-cell absorbers \cite{Shockley61}, excellent absorption properties in the visible part of the solar spectrum \cite{deWolf14}, and high mobilities for both electron and hole transport \cite{Stranks13,Xing13}, as well as low-temperature processing, low manufacturing cost, light weight, and environmental friendliness. Therefore, \text{HPPVs} have become promising candidates for solar-cell devices which can offer clean, affordable and sustainable energy.

Theoretical investigations play an important role in understanding the materials properties and fundamental processes for emergent solar-cell research. Along with the rapid development of HPPV technologies, various theoretical studies have been carried out for hybrid-perovskite-based systems in recent years \cite{Mosconi13,Brivio13,Filippetti14,Castelli14,Yin15}. Interesting aspects of these studies include the crystal structures of hybrid perovskites in different phases \cite{Borriello08,Amat14,Mosconi14,HuangL14}, the importance of spin-orbit coupling (SOC) \cite{Even12,Even13,Brivio14,Katan15}, many-body effects \cite{Umari14,Brivio14,Ahmed14}, and ferroelectric effects for the electronic structure of hybrid-perovskite semiconductors \cite{Frost14a,Frost14b,Leguy15}. However, many challenges remain despite this recent progress. For example, the atomic structure of these materials is still riddled with controversies and it is not yet clear which theoretical approach is most suitable. This is a fundamental question in computational materials science and a prerequisite for further theoretical investigations, especially due to the extreme complexity in the structure of hybrid perovskites. In this context, Brivio \textit{et al.} observed three local total-energy minima associated with different $\text{MA}^+$-orientations in $\text{MAPbI}_3^{}$ \cite{Brivio13}, for which Frost \textit{et al.} obtained a distribution using \textit{ab initio} molecular dynamics, claiming that the orientation of $\text{C}$--$\text{N}$ bonds along the $[100]$ direction (``face-to-face'') is the most populated \cite{Frost14b}. Egger and Kronik found only one preferred $\text{MA}^+$-orientation and attributed its stability to hydrogen bonding \cite{Egger14}. Motta \textit{et al.} reported two local total-energy minima with a $\sim\!\!20~\text{meV}$-per-unit-cell difference in favor of the ``$[011]$''-orientation which is very close to the structure studied by Egger and Kronik but very different from the ``$[110]$'' (``edge-to-edge'') structure of Brivio \textit{et al.} \cite{Motta15a}. Conversely, the calculations by Yin \textit{et al.} indicated that the ``$[111]$''-orientation (along the diagonal of the unit cell) is energetically more favorable \cite{Yin15}. This plethora of competing structural models is a result of different computational approaches, for example, different density functionals, the use of the long-range van der Waals (vdW) interactions, and full relativistic effects.

In this work, we perform a comprehensive analysis of the atomic structure of hybrid perovskites. We focus on the impact of dispersive interactions, which have only been included in three of the many density-functional-theory (DFT) calculations for HPPVs \cite{WangY14,Egger14,Motta15a}. It was shown that the incorporation of vdW interactions into the DFT calculations results in unit-cell volumes in good agreement with experiments. However, the impact of vdW interactions on the atomic structure (including the relative location of the organic cation and the deformation of the inorganic framework) has not been investigated so far. Employing the Tkatchenko-Scheffler (TS) pairwise dispersion scheme \cite{Tkatchenko09}, Egger and Kronik were able to calculate the vdW interaction energy between each interatomic pair for an optimized $\text{MAPbI}_3^{}$ geometry, and found that the iodine-iodine interaction ($\sim\!\!100~\text{meV}$ per pair) has the largest contribution to the overall dispersive (TS) correction of total energy \cite{Egger14}. Nevertheless, more calculations and analysis are required to understand the impact of vdW interactions on the atomic structure of hybrid perovskites.

In this work, we investigated the prototype hybrid perovskite $\text{MAPbI}_3^{}$ and three isostructural systems, methyl-ammonium tin triiodide ($\text{MASnI}_3^{}$), methylammonium lead trichloride ($\text{MAPbCl}_3^{}$) and caesium lead triiodide ($\text{CsPbI}_3^{}$). These three perovskites can be regarded as close relatives of the prototype $\text{MA}_1^{(\text{I})}\text{Pb}_1^{(\text{II})}\text{I}_3^{(-\text{I})}$, as each of them only differs from it by one component: the monovalent cation ($\text{CsPbI}_3^{}$), the divalent cation ($\text{MASnI}_3^{}$), or the halide ($\text{MAPbCl}_3^{}$). Thus, our study facilitates a systematic analysis of interatomic interactions in metal- and organometal-trihalide perovskites. For example, we can compare the vdW interactions between the organic $\text{MA}^+$ cation and the iodide anions in $\text{MAPbI}_3^{}$ and in $\text{MASnI}_3^{}$, or the $\text{MA}$-$\text{I}$ interaction in $\text{MAPbI}_3^{}$ and its counterpart ($\text{Cs}$-$\text{I}$) in $\text{CsPbI}_3^{}$.

Of particular interest in this paper is the question how the organic and the inorganic component in HPPVs interact with each other. Does the inorganic framework deform on its own and the $\text{MA}^+$ cation then accommodates itself in a particular orientation in the deformed cage? Or does the $\text{MA}^+$ cation force the inorganic framework into a particular deformation? We address this \emph{chicken-and-egg} problem with density-functional theory to gain atomistic insight that will be useful for future HPPV design. In addition, we ask the question: which effect vdW interactions and hydrogen bonds have on the interplay between the organic and the inorganic component in HPPVs?

The remainder of this paper is organized as follows: In Section~\ref{model} we outline the model systems and the computational details of our DFT calculations. Section~\ref{results} presents the lattice constants and atomic structures for all considered systems. We discuss the interplay of the organic cation and the inorganic framework as well as the impact of vdW interactions in detail. Finally, Section~\ref{conclusions} concludes with a summary.

\section{Models and computational details}\label{model}

\begin{figure}[!ht]
\begin{center}
\includegraphics[scale=.7]{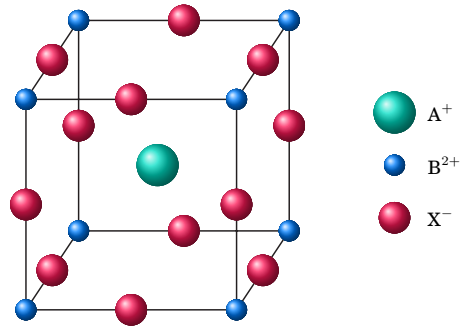}
\end{center}
\caption{Cubic primitive cell of the perovskite structure used to model the investigated compounds $\text{ABX}_3^{}$ in this paper. The monovalent cation $\text{A}^+$, the divalent cation $\text{B}^{2+}$ and the halide anions $\text{X}^-$ are represented by the green, blue and red spheres, respectively.}\label{perovskite}
\end{figure}

We adopted the cubic ($\text{Pm}\bar{3}\text{m}$) primitive-cell model for all compounds thus neglecting the disorder of $\text{MA}^+$ cations. The ions were initially placed at the Wyckoff positions of the $\text{Pm}\bar{3}\text{m}$ space group, as shown in Fig.~\ref{perovskite}: the monovalent cation $\text{A}^+=\text{MA}^+\text{ or Cs}^+$ at 1b, the divalent heavy-metal cation $\text{B}^{2+}=\text{Pb}^{2+}\text{ or Sn}^{2+}$ at 1a, and the halide anions $\text{X}^-=\text{I}^-\text{ or Cl}^-$ at 3d. For hybrid (organic-inorganic) perovskites, the organic $\text{MA}^+$ cation was initially placed at the unit-cell center (the 1b Wyckoff position) with several inequivalent $\text{C}$--$\text{N}$ orientations. This model not only allows for a direct comparison to previous studies, but can also provide a basic and informative description of the atomic structure. We are currently also carrying out first-principles calculations for supercell models. The results will be presented elsewhere in the future.

For each perovskite, we optimized the lattice constant and atomic geometry using the following protocol: (i) The forces on the nuclei were minimized for a fixed shape and size of the cubic unit-cell. (ii) The optimized lattice constant was determined from the minimum of the total-energy \textit{vs.} volume curve. We used the Perdew-Burke-Ernzerhof (PBE) generalized gradient approximation \cite{Perdew96} as the exchange-correlation functional throughout this work. This choice was based on our test calculations, which indicated that hybrid functionals (that combine exact exchange with PBE exchange) change the lattice constants by less than $0.3\%$, but increase the computational cost (both memory and CPU time) by approximately one order of magnitude. SOC, which has already been demonstrated to be very important for the electronic structures of (especially lead-based) hybrid perovskites \cite{Even12,Even13,Brivio14,Katan15}, does not strongly influence the lattice constants as shown by Egger \textit{et al.} \cite{Egger14} and our test calculations. Thus, we only included scalar relativistic effects in our calculations via the zero-order regular approximation \cite{vanLenthe93}. Corrections due to long-range vdW interactions were taken into account by employing the TS method based on the Hirshfeld partitioning of the electron density \cite{Tkatchenko09}. Accordingly, the calculations incorporating TS-vdW are labelled by ``PBE+vdW'' hereafter. We used the default parameters in TS-vdW for neutral atoms (listed in Table~S1 of the Supplemental Material, SM) without explicitly calculating the $C_6^{}$ coefficients based on ionic reference systems \cite{Zhang11,Bucko13,Bucko14}. A $\Gamma$-centered $8\times8\times8$ k-point mesh was used for the periodic DFT calculations. All calculations were performed using the all-electron local-atomic-orbital code \texttt{FHI-aims} \cite{Blum09,HavuV09,Levchenko/etal:2015}.

\section{Results and discussions}\label{results}

\subsection{Lattice constants and orientations of the \texorpdfstring{$\text{C}$}{}--\texorpdfstring{$\text{N}$}{} bond in the optimized geometries}

\begin{figure}[!ht]
\begin{center}
\subfigure[~$\text{MAPbI}_3^{}$-\textbf{a}]{\includegraphics[scale=.7]{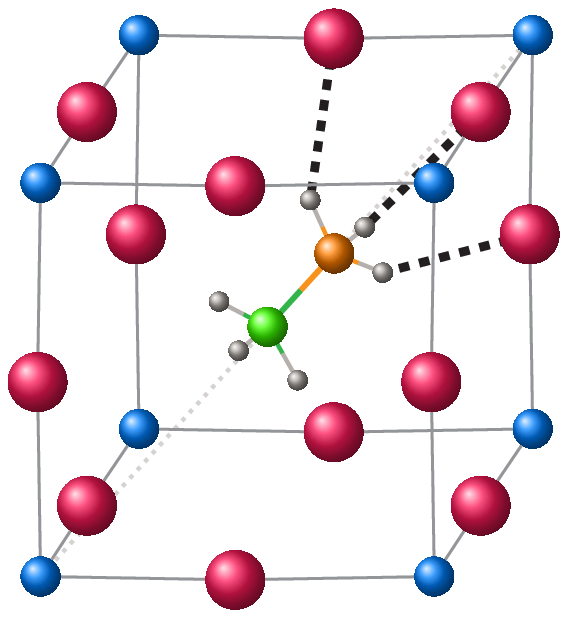}} \quad
\subfigure[~$\text{MAPbI}_3^{}$-\textbf{b}]{\includegraphics[scale=.7]{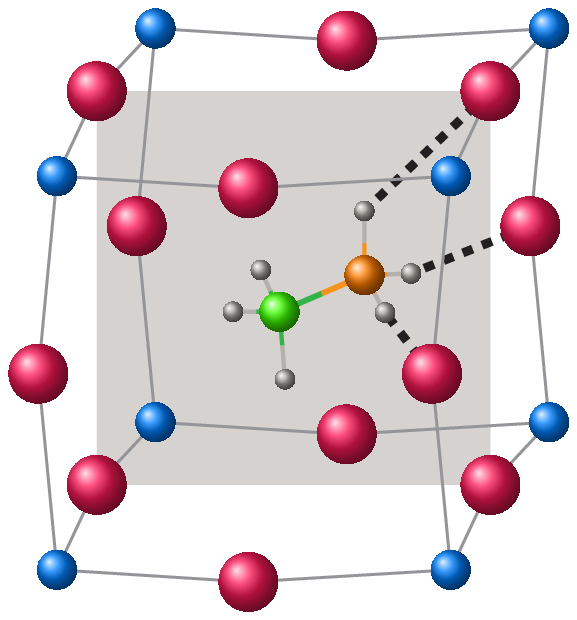}}
\end{center}
\caption{Cubic unit cell of $\text{MAPbI}_3^{}$ optimized with PBE+vdW: (a) structure \textbf{a}, with the colinearity of $\text{Pb}$ (blue), $\text{C}$ (green) and $\text{N}$ (orange) indicated by the gray dotted line; and (b) structure \textbf{b}, with indicating the $(020)$ plane where the $\text{C}$--$\text{N}$ bond is located in gray. The hydrogen bonds between $\text{I}^-$ anions (red) and $\text{H}$ atoms (gray) in the $\text{-NH}_3^+$ group are highlighted by black dashed lines.}\label{PBEvdWstructure}
\end{figure}

At each trial lattice constant, the optimization of our hybrid perovskites results in two local total-energy minima, corresponding to two different structures denoted by $\text{MABX}_3^{}$-\textbf{a} and \textbf{b}, as illustrated in Fig.~\ref{PBEvdWstructure}. For ease of reporting, we shift the whole lattice so that the coordinate of $\text{Pb}^{2+}$ or $\text{Sn}^{2+}$ becomes $(0,0,0)$ (the 1a Wyckoff position). The two geometries correspond to two orientations of the $\text{C}$--$\text{N}$ bond. In case \textbf{a}, the $\text{C}$--$\text{N}$ bond is precisely oriented along the $[111]$ (diagonal) direction and colinear with the unit-cell center as well as the $\text{Pb}^{2+}$ or $\text{Sn}^{2+}$ atom. In case \textbf{b}, the $\text{C}$--$\text{N}$ bond is located in the $(020)$ plane and deviates from the $[100]$ (face-to-face) direction by an angle which depends on the hybrid-perovskite composition, the lattice constant, and the DFT method. For $\text{CsPbI}_3^{}$, the structure remained in the initial cubic geometry during the relaxation.

\begin{figure}[!ht]
\begin{center}
\subfigure[~$\Delta x:\Delta y:\Delta z(\text{C--N})$]{\quad\includegraphics[scale=.7]{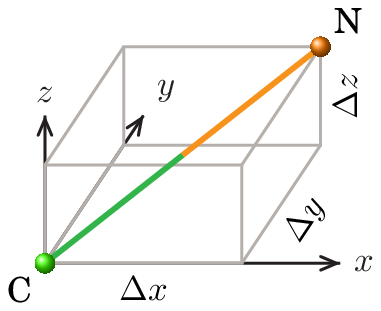}\quad}

\vspace{.5em}
\subfigure[~$u_{\text{NC}}^{}$]{\includegraphics[scale=.7]{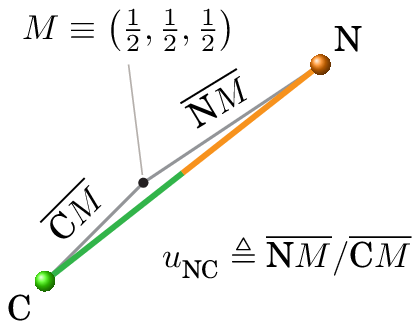}} \quad\quad
\subfigure[~$\varDelta(\text{X})$]{\includegraphics[scale=.7]{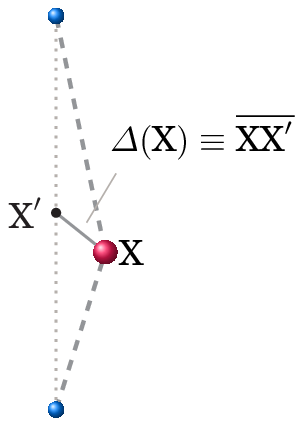}}
\end{center}
\caption{Graphical representation of observables defined to characterize the atomic structure of hybrid perovskites: (a) $\Delta x:\Delta y:\Delta z(\text{C--N})$, (b) $u_{\text{NC}}^{}$ and (c) $\varDelta(\text{X})$. The carbon, nitrogen, halide and divalent metal atoms are colored in green, orange, red and blue, respectively. In (a), $\Delta x$, $\Delta y$ and $\Delta z$ are the projections of the $\text{C}$--$\text{N}$ bond onto the lattice-vector directions $x$, $y$ and $z$, respectively. In (b), $M$ denotes the unit-cell center, and $\overline{\text{N}M}$ and $\overline{\text{C}M}$ are the distances between it and the carbon and nitrogen atoms, respectively. $u_{\text{NC}}^{}$ is defined by their ratio. In (c), $\text{X}^{\prime}$ denotes the Wyckoff position for the $\text{X}^-$ anion, and $\varDelta(\text{X})$ is the distance between it and the position of $\text{X}^-$ in the real system.}\label{structuralquantities}
\end{figure}

For a quantitative description of the atomic structure, we define several quantities that are related to the location of the organic cation and the deformation of the inorganic framework. The definition of these quantities is illustrated graphically in Fig.~\ref{structuralquantities}. The orientation of $\text{MA}^+$ is characterized by $\Delta x:\Delta y:\Delta z(\text{C--N})$, the ratio among the projection of the $\text{C}$--$\text{N}$ bond onto the lattice-vector directions $x$, $y$ and $z$ (Fig.~\ref{structuralquantities}a). The position of $\text{MA}^+$ relative to the center of the cubic unit cell is described by $u_{\text{NC}}^{}$, which is defined by the ratio of the nitrogen-to-cell-center distance to the carbon-to-cell-center distance (Fig.~\ref{structuralquantities}b). The $\varDelta(\text{X})$ values, that is, the deviation of an $\text{X}^-$ cation from its ``ideal'' Wyckoff position in the $\text{Pm}\bar{3}\text{m}$ space group (see Fig.~\ref{structuralquantities}c), is closely related to the deformation of the $\text{BX}_3^-$-framework. Since we are working within the cubic primitive-cell model for all considered systems in this paper, we can use dimensionless fractional coordinates to calculate distances.

The $\Delta x:\Delta y:\Delta z(\text{C--N})$ ratio in case \textbf{a} is $1:1:1$ (or $-1:1:1$, and so forth, for other equivalent structures). In case \textbf{b}, this ratio can be written as $r:0:1$ (or equivalently $-1:r:0$, and so forth) with $r>1$. The angle between the $\text{C}$--$\text{N}$ bond and the $[100]$ direction is $\arctan(r^{-1})$.

Table~\ref{latticeconstants} lists the results of our geometry optimization. First, for each hybrid perovskite, the lattice constant of case \textbf{a} is larger than case \textbf{b}, but the difference is small. Therefore, different orientations of the $\text{C}$--$\text{N}$ bond do not cause large changes in the unit-cell volume in this cubic primitive-cell model. Second, the inclusion of TS-vdW causes a significant reduction of the optimized lattice constants (approximately $2.0\%$, $2.6\%$, $1.6\%$ and $3.7\%$ for $\text{MAPbI}_3^{}$, $\text{MASnI}_3^{}$, $\text{MAPbCl}_3^{}$ and $\text{CsPbI}_3^{}$, respectively). The lattice constants optimized with PBE+vdW agree well with experimental data. Specifically, for both $\text{MAPbI}_3^{}$ and $\text{MASnI}_3^{}$ the overestimation amounts to $\sim\!\!0.6\%$, for $\text{MAPbCl}_3^{}$ the overestimation is $\sim\!\!1.1\%$, and for $\text{CsPbI}_3^{}$ the underestimation is less than $0.2\%$. These lattice constants agree well with the first-principles studies of Egger \textit{et al.} \cite{Egger14} and Motta \textit{et al.} \cite{Motta15a}. Third, structure \textbf{b} is more stable than \textbf{a} for all investigated hybrid compounds as it corresponds to a lower total-energy minimum. A larger total-energy difference (thus also cohesive-energy difference) is obtained when including TS-vdW in the PBE calculations.

\begin{table*}[!ht]
\caption{Results of the geometry optimization (using PBE and PBE+vdW) of the investigated perovskites (in both structures \textbf{a} and \textbf{b} for hybrid systems): lattice constant $a$ of the cubic unit cell (compared with experimental results $a_{\text{exp}}^{}$), orientation of $\text{C}$--$\text{N}$ bond given by the $\Delta x\!:\!\Delta y\!:\!\Delta z(\text{C--N})$ ratio (see Fig.~\ref{structuralquantities}a for definition), and the total-energy difference between structures \textbf{a} and \textbf{b}: $\Delta E_{\text{total}}^{}\triangleq E_{\text{total}}^{}(\text{\textbf{b}})-E_{\text{total}}^{}(\text{\textbf{a}})$. All lattice constants are in $\text{\AA}$; all energies are in $\text{meV}$ (per $\text{ABX}_3^{}$-unit, similarly hereinafter otherwise stated).}\label{latticeconstants}
\begin{center}
\begin{tabular}{ccccccc} \hline\hline
& \quad & \multicolumn{2}{c}{structure \textbf{a}} & \quad & \multicolumn{2}{c}{structure \textbf{b}} \\
&       & ~~~~~~PBE~~~~~~ & ~PBE+vdW~              &       & ~~~~~~PBE~~~~~~ & ~PBE+vdW~              \\ \hline
\multicolumn{7}{l}{$\text{MAPbI}_3^{}$: $a_{\text{exp}}^{}=6.313$ ($>323~\text{K}$) \cite{Stoumpos13}, $6.329$ ($>327~\text{K}$) \cite{Poglitsch87}} \\
$a$                                               & & $6.491$         & $6.364$         & & $6.486$             & $6.350$             \\
$\Delta x\!:\!\Delta y\!:\!\Delta z(\text{C--N})$ & & $1\!:\!1\!:\!1$ & $1\!:\!1\!:\!1$ & & $4.434\!:\!0\!:\!1$ & $2.330\!:\!0\!:\!1$ \\
$\Delta E_{\text{total}}^{}$                      & & --              & --              & & $-10.5$             & $-20.9$             \\ \hline
\multicolumn{7}{l}{$\text{MASnI}_3^{}$: $a_{\text{exp}}^{}=6.231$ ($293~\text{K}$) \cite{Stoumpos13}} \\
$a$                                               & & $6.454$         & $6.268$         & & $6.410$             & $6.257$             \\
$\Delta x\!:\!\Delta y\!:\!\Delta z(\text{C--N})$ & & $1\!:\!1\!:\!1$ & $1\!:\!1\!:\!1$ & & $3.901\!:\!0\!:\!1$ & $2.313\!:\!0\!:\!1$ \\
$\Delta E_{\text{total}}^{}$ & & --               & --              & & ~~$-3.3$            & $-21.6$             \\ \hline
\multicolumn{7}{l}{$\text{MAPbCl}_3^{}$: $a_{\text{exp}}^{}=5.669$ ($280~\text{K}$) \cite{Chi05}} \\
$a$                                               & & $5.843$         & $5.745$         & & $5.811$             & $5.717$             \\
$\Delta x\!:\!\Delta y\!:\!\Delta z(\text{C--N})$ & & $1\!:\!1\!:\!1$ & $1\!:\!1\!:\!1$ & & $3.289\!:\!0\!:\!1$ & $2.483\!:\!0\!:\!1$ \\
$\Delta E_{\text{total}}^{}$ & & --               & --              & & ~~$-7.7$            & $-37.2$             \\ \hline
\multicolumn{4}{l}{$\text{CsPbI}_3^{}$: $a_{\text{exp}}^{}=6.177~\text{\AA}$ ($>583~\text{K}$) \cite{Eperon15}} \\
$a$                                               & & $6.400$         & $6.166$         & & --                  & --                  \\ \hline\hline
\end{tabular}
\end{center}
\end{table*}

We have also performed direct lattice-vector optimization using the analytical stress tensor implemented in \texttt{FHI-aims} \cite{Knuth15}. For case \textbf{a}, the optimized structure remains cubic with slight difference in the lattice constants; while for case \textbf{b}, the cubic symmetry is broken, resulting in an orthorhombic lattice. To focus our analysis, we therefore do not further discuss the results of the stress-tensor optimization in this paper, but will return to it in future work.

Furthermore, we remark that caution has to be applied when comparing DFT lattice constants to experiment. Conventional DFT calculations (for example, the PBE+vdW calculations in this work) are carried out at $0~\text{K}$, while experimental lattice constants for the cubic structures of hybrid perovskites are measured at room temperature or above. For example, the lattice constant $6.313~\text{\AA}$ for $\text{MAPbI}_3^{}$ was measured at above $50~\text{\textdegree C}$ \cite{Stoumpos13}. The lattice constant at $323~\text{K}$ will therefore likely be larger than that at $0~\text{K}$ due to thermal expansion. Preliminary calculations indicate that the thermal expansion is of the order of $0.01~\text{\AA}/100~\text{K}$ \cite{LevardUnpublished} and therefore will only change the lattice constant at $323~\text{K}$ by approximately $0.03~\text{\AA}$.

Table~\ref{structuralparameters} lists the $\text{C}$--$\text{N}$ bond length and the distance between each hydrogen atom in the ammonium group ($\text{H}(\text{N})$) and the nearest halide anion for each optimized geometry. In all cases, the $\text{C}$--$\text{N}$ bond length is $\sim\!\!1.49~\text{\AA}$, and other interatomic distances within the $\text{MA}^+$ cation do not exhibit a pronounced dependence on the composition of the hybrid perovskite or the DFT method (data not shown).

\begin{table}[!ht]
\caption{$\text{C}$--$\text{N}$ bond length $\overline{\text{CN}}$ and hydrogen-bond lengths $\overline{\text{H}(\text{N})\cdots\text{X}}$ (both in $\text{\AA}$) in geometry-optimized perovskites (using PBE and PBE+vdW). Results for both cases \textbf{a} and \textbf{b} are shown.}\label{structuralparameters}
\begin{center}
\begin{tabular}{ccccc} \hline\hline
& \multicolumn{2}{c}{structure \textbf{a}} & \multicolumn{2}{c}{structure \textbf{b}} \\
& ~~~~~~PBE~~~~~~ & ~PBE+vdW~ & ~~~~~~PBE~~~~~~ & ~PBE+vdW~ \\ \hline
\multicolumn{4}{l}{$\text{MAPbI}_3^{}$} \\
$\overline{\text{C--N}}$                           & $1.492$ & $1.490$ & $1.492$ & $1.492$ \\
$\overline{\text{H}(\text{N})\cdots\text{I}}$      & $2.679$ & $2.681$ & $2.656$ & $2.695$ \\
                                                   &         &         & $2.738$ & $2.807$ \\ \hline
\multicolumn{4}{l}{$\text{MASnI}_3^{}$} \\
$\overline{\text{C--N}}$                           & $1.492$ & $1.489$ & $1.492$ & $1.491$ \\
$\overline{\text{H}(\text{N})\cdots\text{I}}$      & $2.661$ & $2.672$ & $2.658$ & $2.677$ \\
                                                   &         &         & $2.719$ & $2.783$ \\ \hline
\multicolumn{4}{l}{$\text{MAPbCl}_3^{}$} \\
$\overline{\text{C--N}}$                           & $1.487$ & $1.485$ & $1.488$ & $1.487$ \\
$\overline{\text{H}(\text{N})\cdots\text{Cl}}$     & $2.273$ & $2.284$ & $2.305$ & $2.310$ \\
                                                   &         &         & $2.338$ & $2.382$ \\\hline\hline
\end{tabular}
\end{center}
\end{table}

The $\text{H}$($\text{N}$)$\cdots\text{X}$ distances indicate typical hydrogen-bonding character for all $\text{MA}$-based compounds. For case \textbf{a}, all three distances are equal to each other, while for case \textbf{b} we obtained two different values. The $\text{H}$($\text{N}$)$\cdots\text{I}$ distances in $\text{MABI}_3^{}$-\textbf{a} and \textbf{b} lie between $2.65$ and $2.81~\text{\AA}$. These values agree well with recent neutron powder diffraction data ($2.61-2.81~\text{\AA}$) for $\text{MAPbI}_3^{}$ \cite{Weller15} and are close to the $\text{H}\cdots\text{I}$ distance ($2.598~\text{\AA}$) in $\text{NH}_4^{}\text{I}$ measured by single crystal neutron diffraction \cite{Seymour70}. They are also very close to the value of $2.65~\text{\AA}$ obtained from a previous \textit{ab initio} Car-Parrinello molecular dynamics study using PBE \cite{Mosconi14}. Similarly, the $\text{H}$($\text{N}$)$\cdots\text{Cl}$ distances in $\text{MAPbCl}_3^{}$ are very close to the value in $\text{NH}_4^{}\text{Cl}$ ($2.32\pm0.02~\text{\AA}$) obtained from neutron-diffraction experiments \cite{Levy52}. This hydrogen-bonding character does not change appreciably when the TS-vdW interaction is not included. We have performed test calculations for $\text{MAPbI}_3^{}$-\textbf{a} in which we switched off the vdW interactions between $\text{H}$ and $\text{I}$. The results show negligible changes in both lattice constant and hydrogen-bond lengths. This demonstrates that PBE describes hydrogen bonding well in hybrid perovskites, which is in line with previous PBE studies of hydrogen bonding \cite{Ireta04,Marom11,Kronik14}.

For $\text{MA}$-based perovskites, structures \textbf{a} and \textbf{b} provide two obvious possibilities for all three hydrogen atoms in the $\text{-NH}_3^+$ group to form hydrogen bonds with halide anions. Other orientations were considered in some previous theoretical works, for example, the $[110]$ (edge-to-edge) orientation \cite{Brivio13,Frost14b}. This orientation is not suitable for hydrogen bonding, because all $\text{N}$--$\text{H}$ bonds point toward face centers or cube corners thus being far away from the halide anions.

\subsection{Atomic structure of the optimized geometries}

We start our discussion of the atomic structure with case \textbf{a}. Table~\ref{a-coordinates} lists some of the coordinates (please refer to Table~S2 for the full data) as well as structural parameters $u_{\text{NC}}^{}$ and $\varDelta(\text{X})$.

\begin{table}[!ht]
\caption{Fractional coordinates of the carbon and nitrogen atoms as well as all inequivalent halide anions in the optimized geometries (using PBE and PBE+vdW) of the investigated hybrid perovskites in structure \textbf{a}. Lead or tin atoms are located at $(0,0,0)$. $\varDelta(\text{X})$ is the deviation of each halide anion from the ideal position: $\big(\frac{1}{2},1,1\big)$, $\big(1,\frac{1}{2},1\big)$ or $\big(1,1,\frac{1}{2}\big)$. Also listed is the $u_{\text{NC}}^{}$ value for each system. All data are dimensionless.}\label{a-coordinates}
\begin{center}
\begin{tabular}{ccccccccccccc} \hline\hline
& $~$ & \multicolumn{5}{c}{PBE} & $~$ & \multicolumn{5}{c}{PBE+vdW} \\
& & $x$ & $y$ & $z$ & $~$ & $\varDelta(\text{X})$ & & $x$ & $y$ & $z$ & $~$ & $\varDelta(\text{X})$ \\ \hline
\multicolumn{13}{l}{$\text{MAPbI}_3^{}$} \\
$\text{C}~~$ & & $0.464$ & $0.464$ & $0.464$ & & & & $0.462$ & $0.462$ & $0.462$ & & \\
$\text{N}~~$ & & $0.596$ & $0.596$ & $0.596$ & & & & $0.597$ & $0.597$ & $0.597$ & & \\
$u_{\text{NC}}^{}$ & & \multicolumn{3}{c}{$2.655$} & & & & \multicolumn{3}{c}{$2.580$} & & \\
$\text{I}~~$ & & $0.487$ & $0.984$ & $0.984$ & & $0.026$ & & $0.496$ & $0.993$ & $0.994$ & & $0.010$ \\
$\text{I}~~$ & & $0.984$ & $0.487$ & $0.984$ & & $0.026$ & & $0.994$ & $0.496$ & $0.993$ & & $0.010$ \\
$\text{I}~~$ & & $0.984$ & $0.984$ & $0.487$ & & $0.026$ & & $0.993$ & $0.994$ & $0.496$ & & $0.010$ \\ \hline
\multicolumn{13}{l}{$\text{MASnI}_3^{}$} \\
$\text{C}~~$ & & $0.461$ & $0.461$ & $0.461$ & & & & $0.452$ & $0.452$ & $0.452$ & & \\
$\text{N}~~$ & & $0.594$ & $0.594$ & $0.594$ & & & & $0.589$ & $0.589$ & $0.589$ & & \\
$u_{\text{NC}}^{}$ & & \multicolumn{3}{c}{$2.417$} & & & & \multicolumn{3}{c}{$1.874$} & & \\
$\text{I}~~$ & & $0.466$ & $0.981$ & $0.981$ & & $0.043$ & & $0.489$ & $0.991$ & $0.991$ & & $0.017$ \\
$\text{I}~~$ & & $0.981$ & $0.466$ & $0.981$ & & $0.043$ & & $0.991$ & $0.489$ & $0.991$ & & $0.017$ \\
$\text{I}~~$ & & $0.981$ & $0.981$ & $0.466$ & & $0.043$ & & $0.991$ & $0.991$ & $0.489$ & & $0.017$ \\ \hline
\multicolumn{13}{l}{$\text{MAPbCl}_3^{}$} \\
$\text{C}~~$ & & $0.452$ & $0.452$ & $0.452$ & & & & $0.443$ & $0.443$ & $0.443$ & & \\
$\text{N}~~$ & & $0.598$ & $0.598$ & $0.598$ & & & & $0.592$ & $0.592$ & $0.592$ & & \\
$u_{\text{NC}}^{}$ & & \multicolumn{3}{c}{$2.031$} & & & & \multicolumn{3}{c}{$1.607$} & & \\
$\text{Cl}~~$ & & $0.484$ & $0.981$ & $0.981$ & & $0.032$ & & $0.492$ & $0.982$ & $0.982$ & & $0.026$ \\
$\text{Cl}~~$ & & $0.981$ & $0.484$ & $0.981$ & & $0.032$ & & $0.982$ & $0.492$ & $0.982$ & & $0.026$ \\
$\text{Cl}~~$ & & $0.981$ & $0.981$ & $0.484$ & & $0.032$ & & $0.982$ & $0.982$ & $0.492$ & & $0.026$ \\ \hline\hline
\end{tabular}
\end{center}
\end{table}

Table~\ref{a-coordinates} shows that all systems of structure \textbf{a} are strictly isotropic in three dimensions due to the rotational symmetry with respect to the three-fold axes along the $[111]$ (diagonal, \textit{cf.} Fig.~\ref{PBEvdWstructure}a) direction. However, the organic cation and all three halide anions are not located at their ideal Wyckoff positions of the $\text{Pm}\bar{3}\text{m}$ space group. The center of $\text{MA}^+$ deviates from the center of the cubic unit cell (the 1b Wyckoff position) so that the nitrogen-end is closer to the $(1,1,1)$-vertex, and the carbon-end moves toward $\big(\frac{1}{2},\frac{1}{2},\frac{1}{2}\big)$ accordingly. Consequently, the three hydrogen atoms of the ammonium group are closer to the three halide anions whose positions are approximately $\big(\frac{1}{2},1,1\big)$, $\big(1,\frac{1}{2},1\big)$ and $\big(1,1,\frac{1}{2}\big)$. This significantly favors the development of $\text{H}$($\text{N}$)$\cdots\text{X}$ distances for hydrogen-bonding. We use the $u_{\text{NC}}^{}$ parameter (Fig.~\ref{structuralquantities}b) to characterize this displacement. Table~\ref{a-coordinates} shows that, by including TS-vdW in the PBE calculations, the $u_{\text{NC}}^{}$ values of all hybrid perovskites decrease, indicating a smaller displacement of $\text{MA}^+$ cation from the unit-cell center. This can be rationalized by simple geometrical considerations based on the $\text{H}$($\text{N}$)$\cdots\text{X}$ distance and the unit-cell-volume reduction caused by the vdW interactions.

The halide anions in hybrid perovskites of structure \textbf{a} are ``pulled'' away from their ideal (3d Wyckoff) positions toward the nitrogen-end of the $\text{MA}^+$ cation. This is reflected by the coordinates of the $\text{X}^-$ anions and their $\varDelta(\text{X})$ values. Such a deformation of the inorganic framework is closely related to the formation of hydrogen bonds. For example, in the PBE geometry of $\text{MAPbI}_3^{}$-\textbf{a}, the distance between the iodide anion at $(0.487,0.984,0.984)$ and a hydrogen atom of the $\text{-NH}_3^+$ group at $(0.512,0.693,0.693)$ (\textit{cf.} Table~S2) is $2.679~\text{\AA}$, while the distance between this $\text{H}$ atom and the corresponding 3d Wyckoff position $\big(\frac{1}{2},1,1\big)$ is $2.821~\text{\AA}$, being $>5\%$ larger than the former value and unlikely appropriate for $\text{H}\cdots\text{I}$ hydrogen bonding. Table~\ref{a-coordinates} reveals that all three inequivalent $\text{X}^-$ anions equally participate in the hydrogen bonding. Therefore, the resulting deviation from their ideal locations exhibits a three-dimensional isotropic character. By incorporating TS-vdW into the DFT calculations, this deviation decreases noticeably especially for both iodide-based perovskites. The deformation of the $\text{BX}_3^-$-framework is very common in materials with perovskite structure. It can trigger several deviations from the ideal cubic structure, such as the distortion of the $\text{BX}_6^{}$-octahedron, the off-center displacement of $\text{B}$ in the octahedron, and the well-known octahedron-tilting \cite{Chi05,Borriello08,Cheng10,Baikie13,Stoumpos13}. To model these features, we would need to go beyond the primitive-cell model in future studies.

\begin{figure}[!ht]
\includegraphics[scale=.64]{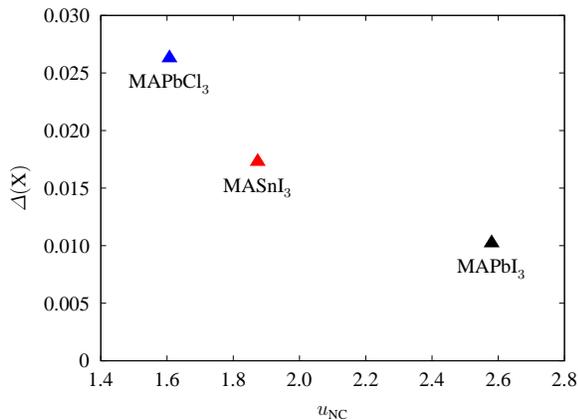}
\caption{PBE+vdW results of $u_{\text{NC}}^{}$ and $\varDelta(\text{X})$ for $\text{MAPbI}_3^{}$-\textbf{a} (black), $\text{MASnI}_3^{}$-\textbf{a} (red) and $\text{MAPbCl}_3^{}$-\textbf{a} (blue).}\label{a-dx_unc}
\end{figure}

From the PBE+vdW results of structure \textbf{a}, we observe an interesting correspondence between $u_{\text{NC}}^{}$ and $\varDelta(\text{X})$ among the three hybrid perovskites: As $u_{\text{NC}}^{}$ decreases from $\text{MAPbI}_3^{}$, $\text{MASnI}_3^{}$ to $\text{MAPbCl}_3^{}$ (so does the lattice constant $a$), that is, the $\text{MA}^+$ cation becomes closer to the unit-cell center, $\varDelta(\text{X})$ increases, indicating a larger magnitude of the inorganic-framework deformation. A graphical representation is given in Fig.~\ref{a-dx_unc}.

Now we turn to structure \textbf{b}. Table~\ref{b-coordinates} lists the coordinates for carbon, nitrogen and halide atoms. We reiterate that there is an essential difference between the geometry of structure \textbf{a} and \textbf{b}: structure \textbf{b} does not show three-dimensional isotropy as structure \textbf{a}, instead it exhibits reflectional symmetry with respect to the $(020)$ plane, in which the $\text{C}$--$\text{N}$ bond is located (\textit{cf.} Fig.~\ref{PBEvdWstructure}b).

\begin{table}[!ht]
\caption{Fractional coordinates of the carbon and nitrogen atoms as well as iodide anions (with the ones involved in hydrogen bonding labelled by superscript ``$^*$'')  in the optimized geometries (using PBE and PBE+vdW) of the investigated hybrid perovskites in structure \textbf{b}. Lead or tin atoms are located at $(0,0,0)$. $\varDelta(\text{X})$ is the deviation of each halide anion from its ideal position: $\big(1,\frac{1}{2},1\big)$, $\big(1,0,\frac{1}{2}\big)$, $\big(1,1,\frac{1}{2}\big)$ or $\big(\frac{1}{2},1,1\big)$ (note: the second and the third are equivalent because of the translational symmetry). Also listed is the $u_{\text{NC}}^{}$ value for each system. All data are dimensionless.}\label{b-coordinates}
\begin{center}
\begin{tabular}{ccccccccccccc} \hline\hline
& $~$ & \multicolumn{5}{c}{PBE} & $~$ & \multicolumn{5}{c}{PBE+vdW} \\
& & $x$ & $y$ & $z$ & $~$ & $\varDelta(\text{X})$ & & $x$ & $y$ & $z$ & $~$ & $\varDelta(\text{X})$ \\ \hline
\multicolumn{13}{l}{$\text{MAPbI}_3^{}$} \\
$\text{C}~~$ & & $0.446$ & $0.500$ & $0.498$ & & & & $0.440$ & $0.500$ & $0.468$ & & \\
$\text{N}~~$ & & $0.670$ & $0.500$ & $0.549$ & & & & $0.656$ & $0.500$ & $0.561$ & & \\
$u_{\text{NC}}^{}$ & & \multicolumn{3}{c}{$3.251$} & & & & \multicolumn{3}{c}{$2.469$} & & \\
$\text{I}^*$ & & $0.956$ & $0.500$ & $1.025$ & & $0.051$ & & $0.976$ & $0.500$ & $1.028$ & & $0.037$ \\
$\text{I}^*$ & & $0.946$ & $0.000$ & $0.495$ & & $0.054$ & & $0.953$ & $0.000$ & $0.498$ & & $0.047$ \\
$\text{I}^*$ & & $0.946$ & $1.000$ & $0.495$ & & $0.054$ & & $0.953$ & $1.000$ & $0.498$ & & $0.047$ \\
$\text{I}~~$ & & $0.472$ & $1.000$ & $0.970$ & & $0.041$ & & $0.486$ & $1.000$ & $0.969$ & & $0.034$ \\ \hline
\multicolumn{13}{l}{$\text{MASnI}_3^{}$} \\
$\text{C}~~$ & & $0.441$ & $0.500$ & $0.481$ & & & & $0.430$ & $0.500$ & $0.464$ & & \\
$\text{N}~~$ & & $0.667$ & $0.500$ & $0.539$ & & & & $0.649$ & $0.500$ & $0.558$ & & \\
$u_{\text{NC}}^{}$ & & \multicolumn{3}{c}{$2.790$} & & & & \multicolumn{3}{c}{$2.034$} & & \\
$\text{I}^*$ & & $0.960$ & $0.500$ & $1.016$ & & $0.043$ & & $0.978$ & $0.500$ & $1.022$ & & $0.032$ \\
$\text{I}^*$ & & $0.953$ & $0.000$ & $0.485$ & & $0.049$ & & $0.957$ & $0.000$ & $0.498$ & & $0.043$ \\
$\text{I}^*$ & & $0.953$ & $1.000$ & $0.485$ & & $0.049$ & & $0.957$ & $1.000$ & $0.498$ & & $0.043$ \\
$\text{I}~~$ & & $0.458$ & $1.000$ & $0.975$ & & $0.049$ & & $0.478$ & $1.000$ & $0.977$ & & $0.032$ \\ \hline
\multicolumn{13}{l}{$\text{MAPbCl}_3^{}$} \\
$\text{C}~~$ & & $0.429$ & $0.500$ & $0.476$ & & & & $0.413$ & $0.500$ & $0.459$ & & \\
$\text{N}~~$ & & $0.674$ & $0.500$ & $0.550$ & & & & $0.654$ & $0.500$ & $0.556$ & & \\
$u_{\text{NC}}^{}$ & & \multicolumn{3}{c}{$2.404$} & & & & \multicolumn{3}{c}{$1.698$} & & \\
$\text{Cl}^*$ & & $0.954$ & $0.500$ & $1.026$ & & $0.053$ & & $0.959$ & $0.500$ & $1.025$ & & $0.048$ \\
$\text{Cl}^*$ & & $0.949$ & $0.000$ & $0.492$ & & $0.052$ & & $0.944$ & $0.000$ & $0.495$ & & $0.057$ \\
$\text{Cl}^*$ & & $0.949$ & $1.000$ & $0.492$ & & $0.052$ & & $0.944$ & $1.000$ & $0.495$ & & $0.057$ \\
$\text{Cl}~~$ & & $0.481$ & $1.000$ & $0.963$ & & $0.042$ & & $0.491$ & $1.000$ & $0.962$ & & $0.039$ \\ \hline\hline
\end{tabular}
\end{center}
\end{table}

We first analyse the geometry of the organic $\text{MA}^+$ cation. This requires two observables: position and orientation. The position of $\text{MA}^+$ is described by the $u_{\text{NC}}^{}$ value characterizing its displacement from the unit-cell center (note: unlike in structure \textbf{a}, the unit-cell center is not colinear with the $\text{C}$--$\text{N}$ bond in structure \textbf{b}). Table~\ref{b-coordinates} shows that, similar to structure \textbf{a}, $u_{\text{NC}}^{}$ decreases when TS-vdW is included in the DFT calculations, which corresponds to a smaller displacement of the $\text{MA}^+$ cation from $\big(\frac{1}{2},\frac{1}{2},\frac{1}{2}\big)$. The $\text{C}$--$\text{N}$ bond in structure \textbf{b} is located within the $(020)$ plane (the $xz$ plane that equally divides the unit cell) with an orientation characterized by the $\Delta x:\Delta y:\Delta z(\text{C--N})$ values listed in Table~\ref{structuralparameters}. For each investigated system, the inclusion of TS-vdW results in a smaller $\Delta x:\Delta z$ ratio, that is, a larger angle between the $\text{C}$--$\text{N}$ bond and the $[100]$ (face-to-face) direction. For example, this angle is $\arctan(4.434^{-1})=12.3\text{\textdegree}$ in the PBE geometry of $\text{MAPbI}_3^{}$, and becomes $\arctan(2.330^{-1})=23.2\text{\textdegree}$ in the PBE+vdW geometry.

Similar to structure \textbf{a}, the position and orientation of $\text{MA}^+$ in structure \textbf{b} adjust to form hydrogen bonds between the hydrogen atoms at the nitrogen-end and the halide anions. Since there are two parameters determining the $\text{MA}^+$-geometry, it is not easy to derive a clear trend among the three investigated hybrid perovskites. Nevertheless, we can still determine some relationships. For example, Table~\ref{structuralparameters} shows very close $\Delta x:\Delta y:\Delta z(\text{C--N})$ ratios for the geometry of the two iodide-based hybrid perovskites optimized with PBE+vdW, while the lattice constant of $\text{MASnI}_3^{}$-\textbf{b} is noticeably smaller than that of $\text{MAPbI}_3^{}$-\textbf{b}. Therefore a larger deviation of $\text{MA}^+$ from the unit-cell center in $\text{MAPbI}_3^{}$-\textbf{b} is required for $\text{H}$($\text{N}$)$\cdots\text{I}$ bonding, as demonstrated by the fact that $u_{\text{NC}}^{}$ of $\text{MAPbI}_3^{}$-\textbf{b} is larger than of $\text{MASnI}_3^{}$-\textbf{b}.

The $\text{X}^-$ anions deviate from their ideal 3d Wyckoff positions as shown in Table~\ref{b-coordinates}. This is closely related to the $\text{H}$($\text{N}$)$\cdots\text{X}$ hydrogen bonding. Because of the primitive-cell model and the reflectional symmetry with respect to the $(020)$ plane, this deviation is limited to the $xz$ plane. Different to structure \textbf{a} where all halide anions equally participate in the hydrogen bonding, the three halides in the unit cell of structure \textbf{b} play different roles when interacting with the $\text{-NH}_3^+$ group. Specifically, the $\text{X}^-$ anion approximately located at $\big(1,\frac{1}{2},1\big)$ (the first $\text{X}^*$ in each block of Table~\ref{b-coordinates}, which is located in the $(020)$ plane) forms a hydrogen bond with one hydrogen of the $\text{-NH}_3^+$ group. In PBE geometries, this halide anion is significantly ``pulled'' toward the $\text{MA}^+$ cation along $x$, and ``pushed'' away from the $\text{MA}^+$ cation along $z$ (that is, the $z$-coordinate is larger than one). The deviation along $x$ is larger than along $z$ by a factor of $\sim\!\!2$. For both $\text{MAPbI}_3^{}$-\textbf{b} and $\text{MASnI}_3^{}$-\textbf{b}, the inclusion of TS-vdW reduces its $x$-deviation by a factor of $\sim\!\!2$, and slightly enlarges the $z$-deviation. This is an accumulative result of the $u_{\text{NC}}^{}$-decrease and the reduced $\Delta x:\Delta z(\text{C--N})$ ratio: they cause a decrease of the $x$-coordinate together with an increase of $z$ of the nitrogen's position, and accordingly a movement of this halide's position.

Two $\text{X}^-$ anions, which are approximately located at $\big(1,0,\frac{1}{2}\big)$ and $\big(1,1,\frac{1}{2}\big)$, and therefore are equivalent to each other, form hydrogen bonds with the other two hydrogen atoms at the nitrogen-end. Each of them simultaneously forms two hydrogen bonds with $\text{MA}^+$ cations in neighbouring unit cells because of the periodic boundary condition. Table~\ref{b-coordinates} shows that, the deviation of these halides' positions from the 3d Wyckoff positions along $x$ is much larger than that along $z$, and the effect of TS-vdW on their positions is rather small. Besides, the $\text{X}^-$ anion approximately located at $\big(\frac{1}{2},1,1\big)$ or equivalent positions (the one without the superscript-label ``$^*$'' in Table~\ref{b-coordinates}) does not participate in hydrogen bonding. Nevertheless, its deviation along $z$ is significantly larger than each of the other two halide anions, and this deviation is nearly unaffected by the inclusion of TS-vdW. In addition, TS-vdW reduces its deviation along $x$ by a factor of $\sim\!\!2$. Moreover, the impact of TS-vdW on the the positions of the chloride anions in $\text{MAPbCl}_3^{}$ is much smaller than that for both iodide-based hybrid compounds, which agrees with the fact that the vdW interactions cause the smallest change of $\Delta x:\Delta z(\text{C--N})$ in this system.

\begin{figure}[!ht]
\includegraphics[scale=.64]{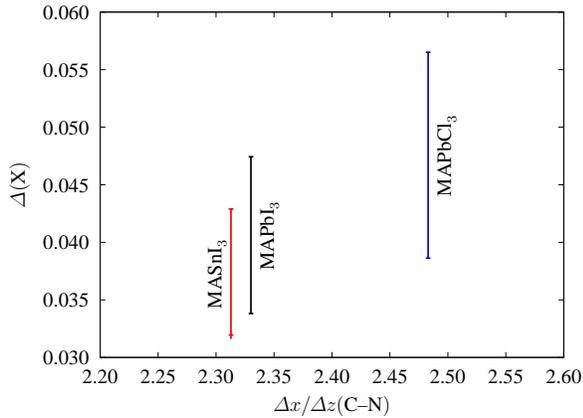}
\caption{PBE+vdW results of $\Delta x/\Delta z(\text{C--N})$ and the $\varDelta(\text{X})$-range for $\text{MAPbI}_3^{}$-\textbf{b} (black), $\text{MASnI}_3^{}$-\textbf{b} (red) and $\text{MAPbCl}_3^{}$-\textbf{b} (blue).}\label{b-dx_xz}.
\end{figure}

From the PBE+vdW results of structure \textbf{b}, we observe a positive correlation between the $\Delta x/\Delta z(\text{C--N})$-increase and the increase of the $\varDelta(\text{X})$-range from $\text{MASnI}_3^{}$, $\text{MAPbI}_3^{}$ to $\text{MAPbCl}_3^{}$. A graphical representation is given in Fig.~\ref{b-dx_xz}.

\subsection{The deformation of the inorganic framework and its interplay with the organic cation \---- a chicken-and-egg paradox}

Next we analyse the atomic structure in more detail to decipher the interplay between the organic cation and the inorganic-framework. We will also investigate the role of vdW interactions in this interplay.

The $\varDelta(\text{X})$ values in Tables~\ref{a-coordinates} and \ref{b-coordinates} allow us to compare the inorganic-framework deformation between structures \textbf{a} and \textbf{b}. For both $\text{MAPbI}_3^{}$ and $\text{MAPbCl}_3^{}$ calculated with PBE, the $\text{PbX}_3^-$-deformation in structure \textbf{b} is larger than in structure \textbf{a} by a factor of $\sim\!\!2$, while for $\text{MASnI}_3^{}$ the $\text{SnI}_3^-$-deformation in \textbf{b} is only slightly larger than in \textbf{a}. The inclusion of TS-vdW in the PBE calculations causes a $\sim\!\!60\%$ reduction of $\varDelta(\text{I})$ in both $\text{MAPbI}_3^{}$-\textbf{a} and $\text{MASnI}_3^{}$-\textbf{a}, and a less than $25\%$ reduction in the corresponding \textbf{b} structures. As a result, the $\text{PbI}_3^-$-deformation in the PBE+vdW structure of $\text{MAPbI}_3^{}$-\textbf{b} is larger than in $\text{MAPbI}_3^{}$-\textbf{a} by a factor of $\sim\!\!4$ (this is intuitively reflected by the much more pronounced $\text{Pb}$--$\text{I}$--$\text{Pb}$ bending in Fig.~\ref{PBEvdWstructure}b \textit{vs.} Fig.~\ref{PBEvdWstructure}a), while the $\text{SnI}_3^-$-deformation in the PBE+vdW structure of $\text{MASnI}_3^{}$-\textbf{b} is larger than in $\text{MASnI}_3^{}$-\textbf{a} by a factor of $\sim\!\!2$. The inclusion of vdW interactions causes a $\sim\!\!20\%$ reduction of $\varDelta(\text{Cl})$ in $\text{MaPbCl}_3^{}$-\textbf{a}, while the impact on $\text{MaPbCl}_3^{}$-\textbf{b} is weak on average.

The trend of the $\text{BX}_3^-$-deformation qualitatively agrees with the trend of the cohesive-energy difference ($\Delta E_{\text{total}}^{}$) listed in Table~\ref{structuralparameters}. For example, in PBE, the cohesive-energy difference between $\text{MASnI}_3^{}$-\textbf{a} and \textbf{b} is very small ($3.3~\text{meV}$), corresponding to the similar magnitude of the $\text{SnI}_3^-$-deformation in these two structures. In contrast, the cohesive-energy difference in PBE+vdW becomes larger by one order of magnitude ($21.6~\text{meV}$), corresponding to the apparently larger $\text{SnI}_3^-$-deformation in $\text{MASnI}_3^{}$-\textbf{b} than in \textbf{a}.

\begin{figure*}[!ht]
\begin{center}
\includegraphics{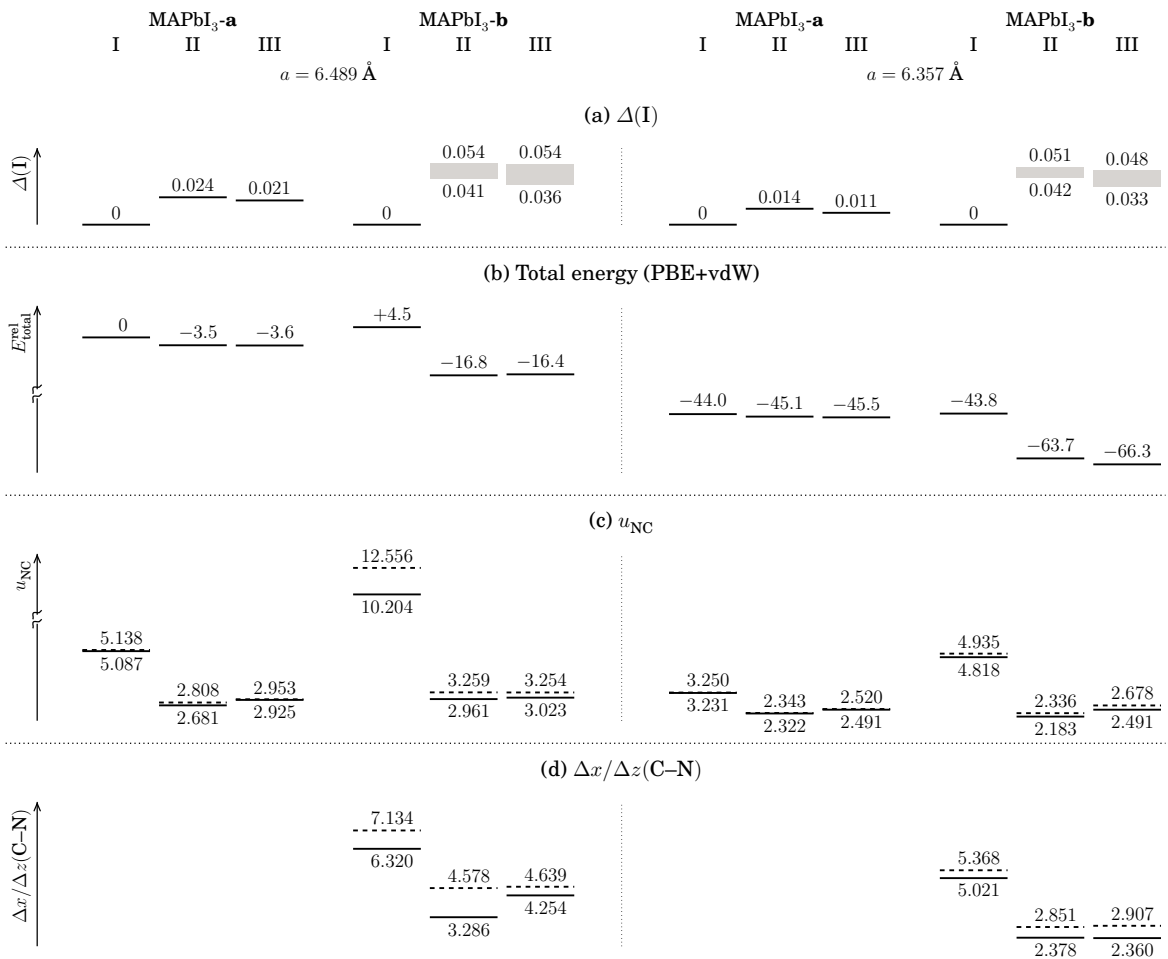}
\end{center}
\caption{DFT results of $\text{MAPbI}_3^{}$-\textbf{a} and \textbf{b} in models I, II and III. Shown are: (a) the $\varDelta(\text{I})$ values (or ranges for \textbf{b}-II and \textbf{b}-III) which define the models, (b) the PBE+vdW total energies, (c) $u_{\text{NC}}^{}$, and (d) $\Delta x/\Delta z(\text{C--N})$ for structure \textbf{b}. Results calculated at both PBE and PBE+vdW lattice constants ($6.489$ and $6.357~\text{\AA}$, respectively) are shown. In (c) and (d), PBE and PBE+vdW results are given in dashed and solid lines, respectively.}\label{MPIscheme}
\end{figure*}

For a systematic analysis, we carried out PBE and PBE+vdW calculations for three models of $\text{MAPbI}_3^{}$-\textbf{a} and \textbf{b}. Our aim is to disentangle the direct effect of the vdW interactions from the indirect effect they have on the lattice constant. We start from an ideal $\text{PbI}_3^-$-framework in model I, that is, no inorganic-framework deformations. We evaluate model I at both the PBE and the PBE+vdW lattice constants. Then in model II, we let the atomic positions relax in PBE at these two fixed lattice constants. Model II thus measures the effect of the distortion of the inorganic framework for the two different molecular orientations \textbf{a} and \textbf{b} as well as the effect of the lattice constant. In model III we then switch on the vdW interactions at the two lattice constants and relax the structures again. This model measures the effect of the vdW interactions at fixed lattice constant. Figure~\ref{MPIscheme} reports the structural-parameters and (PBE+vdW) total-energies for the three models (we do not report the PBE total energies to simplify our discussions).

Figures~\ref{MPIscheme}a and b reveal that the inorganic-framework deformation has the largest effect. Going from model I to II decreases the total energy (and thus increases the stability) while simultaneously $\varDelta(\text{I})$ increases, that is, the inorganic framework deforms. This effect is more pronounced for structure \textbf{b}. This effect is fairly insensitive to the lattice constant.

Figures~\ref{MPIscheme}a and b further reveal that direct vdW effects are small. Going from model II to III only results in a minute energy gain (or even increase at the PBE lattice constant). The corresponding inorganic-framework distortion reduces slightly, manifested in a decrease of $\varDelta(\text{I})$. This is due to the smaller $\text{Pb}$--$\text{I}$ distances in PBE+vdW.

Figures~\ref{MPIscheme}c and d illustrate how the lattice constant, the $\text{PbI}_3^-$-deformation and vdW interactions influence the position of $\text{MA}^+$. $u_{\text{NC}}^{}$ is very sensitive to $\varDelta(\text{I})$: as $\varDelta(\text{I})$ increases from I, III to II, $u_{\text{NC}}^{}$ decreases, indicating that the $\text{MA}^+$ cation moves closer to the unit-cell center. These findings agree with those of Fig.~\ref{a-dx_unc}, where we also observed an inverse relation between $\varDelta(\text{I})$ and $u_{\text{NC}}^{}$ for different hybrid perovskites. Moreover, the impact of vdW interactions on $u_{\text{NC}}^{}$ is generally very small. Similar trends are found for structure \textbf{b} (Fig.~\ref{MPIscheme}d): When the $\text{PbI}_3^{2-}$-deformation is switched on by going from model I to II, $\Delta x/\Delta z(\text{C--N})$ significantly decreases, indicating a larger angle between the $\text{C}$--$\text{N}$ bond and the $[100]$ direction. The effects of vdW interactions are again not large.

Recapitulating: a larger deformation of the inorganic framework corresponds to a lower total energy in $\text{MAPbI}_3^{}$. A reasonable hypothesis is that the deformation of the inorganic $\text{BX}_3^-$-framework of hybrid perovskites is energetically favorable. Structure \textbf{b} is noticeably more stable, because our PBE+vdW results indicate that the orientation of the organic $\text{MA}^+$ cation allows for a larger $\text{BX}_3^-$-deformation.

Can we then postulate that the $\text{BX}_3^-$-deformation intrinsically exists to stabilize the trihalide perovskites? We have carried out test calculations (PBE+vdW) for the reference compound $\text{CsPbI}_3^{}$ using different models (for example, the stress-tensor optimization of supercells). The results indicate that the ideal cubic perovskite structure without $\text{PbI}_6^{}$-deformation is the most stable, which would invalidate our postulate.

\begin{table*}[!ht]
\caption{Dependence of DFT results for $\text{MAPbI}_3^{}$ ($\text{MA}^+$-location, lattice constant, $\text{PbI}_3^-$-deformation and PBE+vdW total energy) on the $\text{MA}^+$-orientation, computational environments (lattice constant and $\text{PbI}_3^-$-deformation) and DFT method (vdW interactions).}\label{MPIsumm}
\begin{center}
\begin{tabular}{lcccccccc} \hline\hline
\multicolumn{1}{c}{DFT results} & \quad & $\text{MA}^+$-orientation & \quad & lattice constant & \quad & $\text{PbI}_3^-$-deformation & \quad & vdW interactions \\ \hline
$\text{MA}^+$-location          & &                --               & & strong                 & & strong                             & & weak                   \\
lattice constant                & & weak                            & &           --           & & weak                               & & strong                 \\
$\text{PbI}_3^-$-deformation    & & strong                          & & moderate               & &              --                    & & moderate               \\
PBE+vdW total energy            & & weak                            & & strong                 & & strong                             & &        --                \\
\hline\hline
\end{tabular}
\end{center}
\end{table*}

The major difference between $\text{CsPbI}_3^{}$ and the $\text{MA}^+$-based hybrid perovskites is the symmetry of the monovalent cation: $\text{Cs}^+$ is spherical, while $\text{MA}^+$ belongs to $\text{C}_{3\text{v}}^{}$. Accordingly, the primitive-cell symmetry of $\text{CsPbI}_3^{}$, $\text{MAPbI}_3^{}$-\textbf{a} and $\text{MAPbI}_3^{}$-\textbf{b} descends as $\text{O}_{\text{h}}^{}\to\text{C}_{3\text{v}}^{}\to\text{C}_{\text{s}}^{}$. Our results show that the $\text{BX}_3^-$-deformation increases as the symmetry reduces. To uncover the physics behind this relation, further calculations and analysis would be required. However, the trend is supported by a recent theoretical study of a similar system, $\text{CsSnI}_3^{}$ \cite{Patrick15}, in which the $\text{SnI}_3^-$-deformation is closely related to the phonon modes. The motion of the phonon can introduce an instantaneous symmetry break-down. However, the phonon motion is temperature-dependent and its modelling would go beyond the DFT methods employed in this article.

To summarize, the organic and the inorganic components of hybrid perovskites act synergetically: The low symmetry of the organic cation (the \emph{chicken}) triggers the inorganic-framework (the \emph{egg}) deformation, whose magnitude is sensitive to the orientation of the organic cation; this deformation then aids the overall stabilization of the hybrid perovskite structure. The final location of the organic cation depends sensitively on the inorganic-framework deformation. We are thus left with a chicken-and-egg paradox that makes it hard to say which came first, the deformation of the inorganic-framework or position of the organic cation.

Table~\ref{MPIsumm} sums up how the DFT total energy and structural parameters of $\text{MAPbI}_3^{}$ depend on different factors. Only \emph{direct} dependences are listed. For example, the location of $\text{MA}^+$ ($u_{\text{NC}}^{}$ and/or $\Delta x/\Delta z(\text{C--N})$) sensitively depends on the lattice constant, which is significantly reduced by vdW interactions. However, when all other factors (lattice constant and the $\text{PbI}_3^-$-deformation) are fixed, vdW interactions have only a very weak effect on the location of $\text{MA}^+$.

\subsection{Impact of vdW interactions on the lattice constants}

Our analysis of the hybrid-perovskite geometries have revealed the importance of incorporating vdW interactions into DFT calculations. The impact of vdW interactions is mainly \emph{indirect}, as the change of the atomic structure is mainly correlated with the unit-cell volume. Next we analyse which component of the hybrid perovskites has the largest influence on the lattice constants. The TS pairwise interatomic scheme implemented in \texttt{FHI-aims} allows us to switch individual vdW interactions between atom pairs on or off. For example, for $\text{MAPbI}_3^{}$-\textbf{a}, we start from PBE reference calculations, and switch on the TS-vdW interaction selectively between different pairs, such as the $\text{MA}^+$ cations (denoted $\text{MA}$-$\text{MA}$), $\text{Pb}$-$\text{Pb}$, $\text{I}$-$\text{I}$, $\text{MA}$-$\text{Pb}$, $\text{MA}$-$\text{I}$, or $\text{Pb}$-$\text{I}$. For simplicity, we regard the $\text{MA}^+$ cation as a whole ``particle'', that is, for $\text{MA}$-$\text{MA}$ calculations, we switch on the TS-vdW interactions $\text{C}$-$\text{C}$, $\text{N}$-$\text{N}$, $\text{H}$-$\text{H}$, $\text{C}$-$\text{N}$, $\text{C}$-$\text{H}$, and $\text{N}$-$\text{H}$; all these interactions are switched off for other calculations, for example, $\text{MA}$-$\text{I}$. This would introduce some error, as the vdW interactions within the same $\text{MA}^+$ cation is considered when calculating $\text{MA}$-$\text{MA}$, but neglected for other calculations. We expect this error to be small, since the internal structure of $\text{MA}^+$ does not depend sensitively on the chosen DFT method as alluded to earlier.

\begin{figure}[!ht]
\begin{center}
\includegraphics[scale=.64]{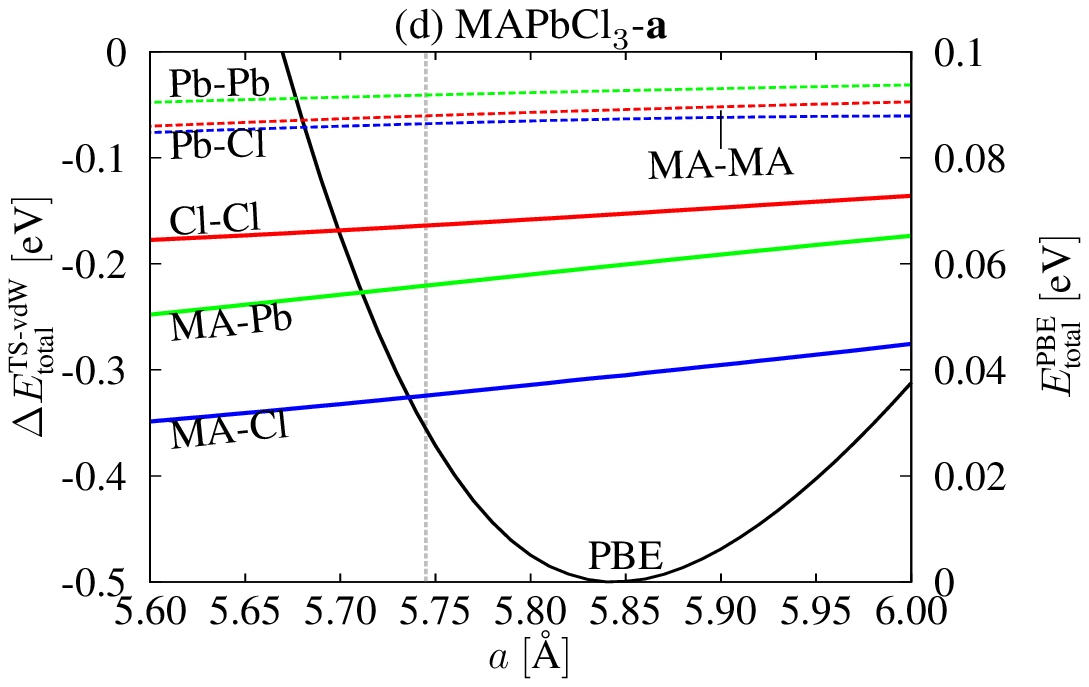}

\vspace{1.em}
\includegraphics[scale=.64]{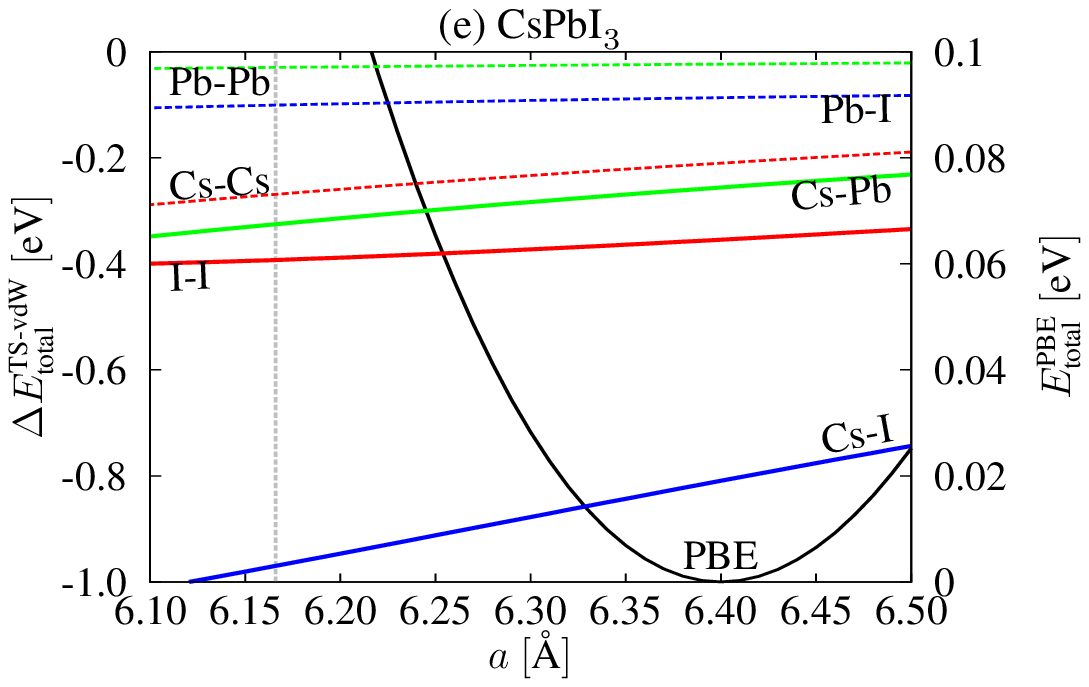}
\end{center}
\vspace{4.em}
\caption{TS-vdW contributions for different interatomic pairs $\Delta E_{\text{total}}^{\text{TS-vdW}}$ for several perovskites: (a) $\text{MAPbI}_3^{}$-\textbf{a}, (b) $\text{MAPbI}_3^{}$-\textbf{b}, (c) $\text{MASnI}_3^{}$-\textbf{a}, (d) $\text{MAPbCl}_3^{}$-\textbf{a} and (e) $\text{CsPbI}_3^{}$ (note the different scale for (e)). Also shown are the PBE total energy curves vertically shifted so that the minimal energy lies at $0$ (black curves, note the different scale of $E_{\text{total}}^{\text{PBE}}$ axes). The lattice constants optimized with PBE+vdW are indicated by the vertical gray dashed lines.}\label{vdWcorrection}
\vspace{1.em}
\end{figure}

We have performed our analysis by scanning over a certain lattice-constant range. At each lattice constant, we optimize the geometry with PBE. The total-energy corrections for the pairwise TS-vdW interactions are plotted in Fig.~\ref{vdWcorrection}. When analysing the impact of vdW on the lattice constant, the meaningful observable is not $\Delta E_{\text{total}}^{\text{TS-vdW}}$, which is defined by the total-energy difference between the PBE+vdW and the PBE calculations, but rather its gradient with respect to the lattice constant $a$.

Our calculations semi-quantitatively reproduce the results of Egger and Kronik \cite{Egger14} in that the vdW interaction between the iodine atoms provides the largest interatomic contribution ($\sim\!\!100~\text{meV}$ per pair) in $\text{MAPbI}_3^{}$ (Fig.~\ref{vdWcorrection}a and b), while in $\text{MAPbCl}_3^{}$ (Fig.~\ref{vdWcorrection}d) the inter-halide interaction energy decreases by a factor of $\sim\!\!2$. However, Egger and Kronik have not summed up the interaction involving the carbon, nitrogen and hydrogen atoms, whichs play an important role in the total-energy correction as shown in Fig.~\ref{vdWcorrection}. They have also not carried out an analysis for different lattice constants, and therefore could not determine the contribution of different pairs to the lattice constant.

Among the convex $E_{\text{total}}^{\text{PBE}}$ quasi-parabolas, the curve for $\text{MASnI}_3^{}$-\textbf{a} (Fig.~\ref{vdWcorrection}c) is the widest, while the curve of $\text{MAPbCl}_3^{}$-\textbf{a} (Fig.~\ref{vdWcorrection}d: only for $a<5.85~\text{\AA}$, that is, the ``left half'' of the curve) is the narrowest. The total-energy correction for each considered vdW pair becomes more negative for decreasing $a$. Hence, as expected, the inclusion of vdW interactions leads to a smaller unit-cell volume. We also observe that (i) most of the $\Delta E_{\text{total}}^{\text{TS-vdW}}$ corrections are approximately linear in $a$, and (ii) in different perovskites, the same pair has roughly the same effect on the lattice constant. For example, the $\text{MA}$-$\text{I}$ curves for $\text{MAPbI}_3^{}$-\textbf{a} (Fig.~\ref{vdWcorrection}a), $\text{MAPbI}_3^{}$-\textbf{b} (Fig.~\ref{vdWcorrection}b) and $\text{MASnI}_3^{}$-\textbf{a} (Fig.~\ref{vdWcorrection}c) agree to within $2.5~\text{meV}$ in the interval $a\in[6.20,6.55]$.

\begin{figure}[!ht]
\begin{center}
\includegraphics[scale=.64]{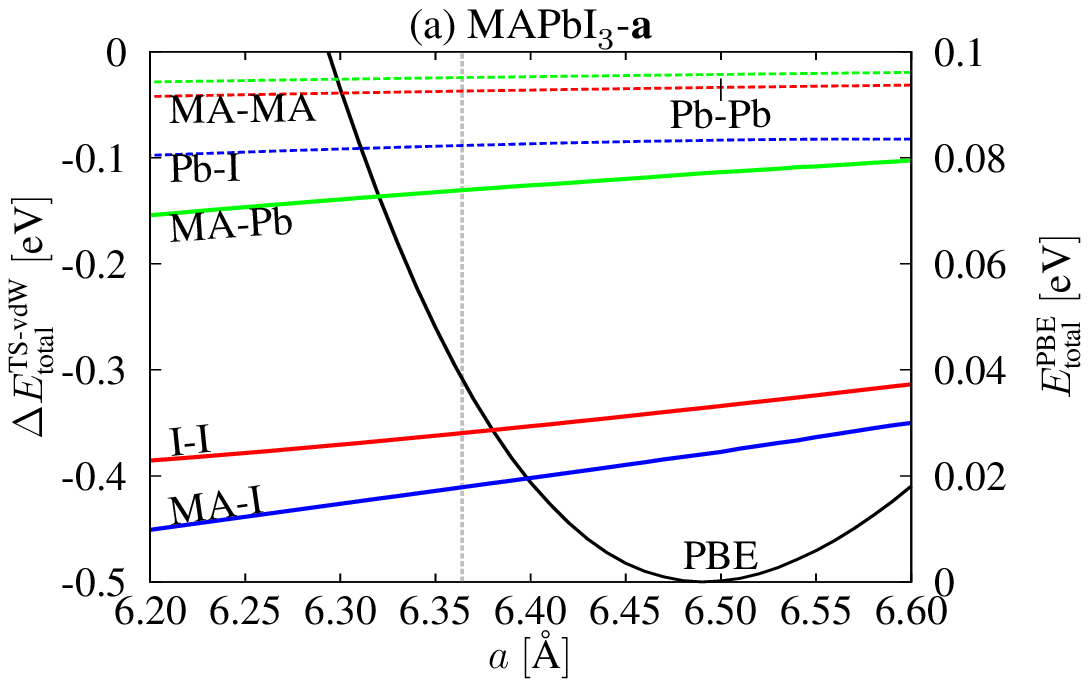}

\vspace{1.em}
\includegraphics[scale=.64]{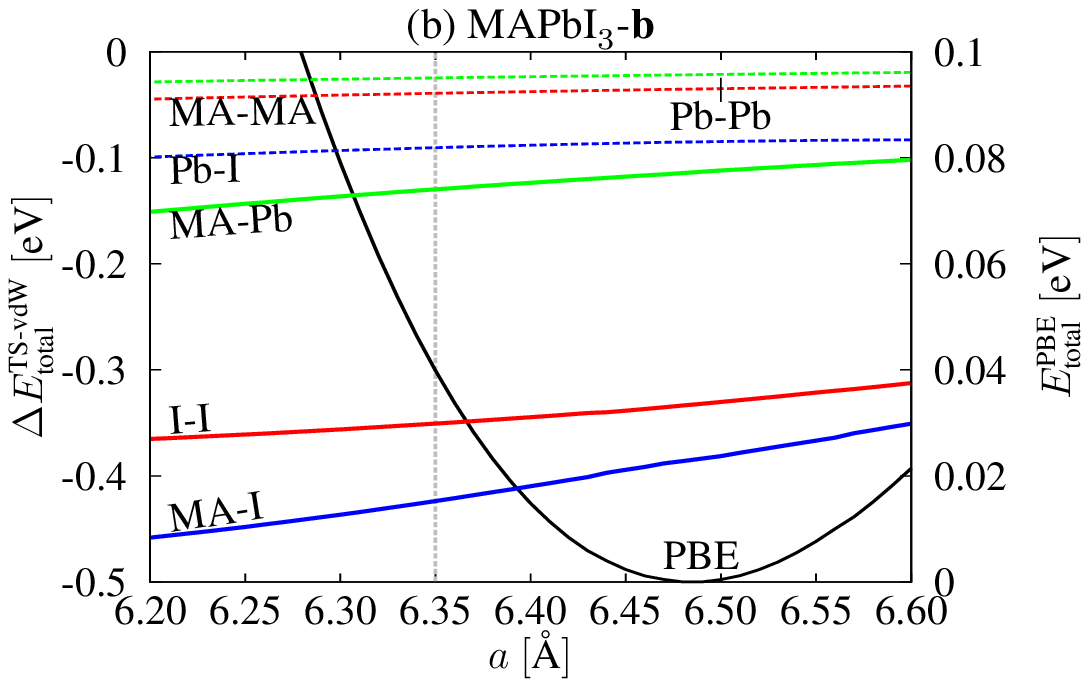}

\vspace{1.em}
\includegraphics[scale=.64]{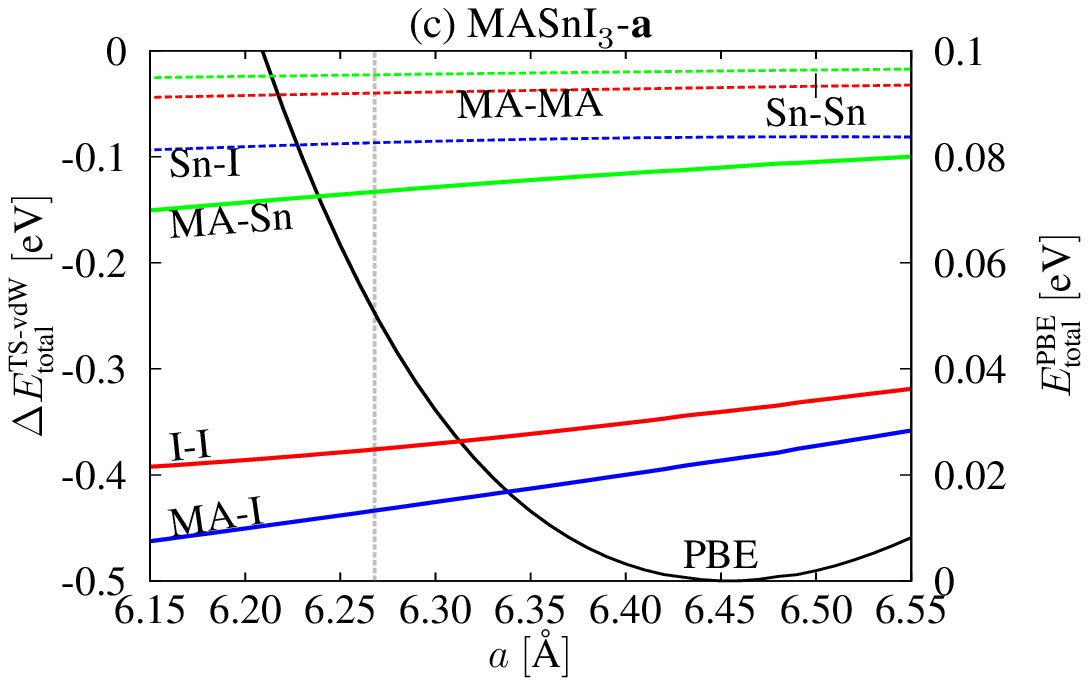}
\end{center}
\end{figure}

For a quantitative analysis, we have performed a second-order polynomial fit (using the nonlinear least-squares Levenberg-Marquardt algorithm) to the PBE total-energy data
\begin{align}
E_{\text{total}}^{\text{PBE}} &= \alpha^{\text{PBE}} a^2 + \beta^{\text{PBE}} a + \gamma^{\text{PBE}} \label{PBEtotal}
\end{align}
and linear fits to the vdW total-energy corrections
\begin{align}
\Delta E_{\text{total}}^{\text{TS-vdW}}(\text{Y-Z}) &= \beta^{\text{TS-vdW}}(\text{Y-Z}) a + \gamma^{\text{TS-vdW}}(\text{Y-Z}) \label{TSvdWcorr}
\end{align}
for each system, with $\text{Y}$-$\text{Z}$ labelling all considered inter-particle pairs. The optimized lattice constant $a^*$ calculated purely with PBE is given by the location of the minimum of the fitted second-order polynomial (\ref{PBEtotal}), while the one including the $\text{Y}$-$\text{Z}$ interaction is given by the minimum-location of the second-order polynomial combining Eqs.~(\ref{PBEtotal}) and (\ref{TSvdWcorr}):
\begin{align}
a^*_{\text{PBE}} &= -\frac{\beta^{\text{PBE}}}{2\alpha^{\text{PBE}}}, \label{PBEa} \\
a^*_{\text{PBE+vdW}}(\text{Y-Z}) &= -\frac{\beta^{\text{PBE}}+\beta^{\text{TS-vdW}}(\text{Y-Z})}{2\alpha^{\text{PBE}}}. \label{vdWa}
\end{align}

Table~\ref{vdWpairwise} shows the results for $\text{MAPbI}_3^{}$-\textbf{b} and $\text{CsPbI}_3^{}$ (the results for all hybrid perovskites in structure \textbf{a} are given in Table~S3). We have omitted the $\Delta E_{\text{total}}^{\text{TS-vdW}}(\text{Pb-I})$-data with $a>6.50~\text{\AA}$ of $\text{MAPbI}_3^{}$-\textbf{b} for a better linear fit. This is safe, as these data do not play any role in the lattice-constant reduction. For $\text{MAPbI}_3^{}$-\textbf{b}, Egger and Kronik identified the $\text{I}$-$\text{I}$ interaction to be most dominant \cite{Egger14}. However, Fig.~\ref{vdWcorrection} reveals that the $\text{MA}$-$\text{I}$ interaction is even larger. Furthermore, the gradient of the $\text{MA}$-$\text{I}$ / $\text{MA}$-$\text{Pb}$ line is steeper than / similar to the $\text{I}$-$\text{I}$ line, which implies that the $\text{MA}$-$\text{I}$ / $\text{MA}$-$\text{Pb}$ interaction has a larger / similar influence on the lattice constant. Finally, $\text{MA}$-$\text{MA}$ and $\text{Pb}$-$\text{Pb}$ contribute only little (less than $1\text{\textperthousand}$). This might be due to the very large inter-particle distances.

\begin{table}[!ht]
\caption{Effects of the TS-vdW interactions for each atom or particle pair for $\text{MAPbI}_3^{}$-\textbf{b} and $\text{CsPbI}_3^{}$. For each system, we fit the second-order polynomial $E_{\text{total}}^{\text{PBE}}=\alpha^{\text{PBE}}a^2+\beta^{\text{PBE}}a+\gamma^{\text{PBE}}$ to the total energy calculated with PBE and list $\alpha^{\text{PBE}}$ and $\beta^{\text{PBE}}$. We also fit the linear function $\Delta E_{\text{total}}^{\text{TS-vdW}}=\beta^{\text{TS-vdW}}a+\gamma^{\text{TS-vdW}}$ to each vdW total-energy correction. Also listed are the optimized lattice constant $a^*$ (calculated using Eq.~(\ref{PBEa}) or (\ref{vdWa})) and its deviation $\Delta a^*$ to the PBE result. For the PBE+vdW results, all listed $\Delta a^*$-deviations are summed up to a value shown in the ``Sum'' row.}\label{vdWpairwise}
\begin{center}
\begin{tabular}{lcccccc} \hline\hline
& $~$ & $a^*~[\text{\AA}]$ & $\Delta a^*~[\text{\AA}]$ & $~$ & \multicolumn{2}{c}{Fitting parameters} \\ \hline
\multicolumn{7}{l}{$\text{MAPbI}_3^{}$-\textbf{b}} \\
PBE                          & &         &                                         & & $\alpha^{\text{PBE}}$ & $\beta^{\text{PBE}}$ \\
                             & & $6.486$ &                                         & & $2.4970$ & $-32.392$ \\
PBE+vdW                      & &         &                                         & & & $\beta^{\text{TS-vdW}}$ \\
with $\text{MA}$-$\text{MA}$ & & $6.480$ & $-0.006~~(-0.9\text{\textperthousand})$ & & & $~~~\,\,\,0.031$ \\
with $\text{Pb}$-$\text{Pb}$ & & $6.482$ & $-0.004~~(-0.7\text{\textperthousand})$ & & & $~~~\,\,\,0.022$ \\
with $\text{I}$-$\text{I}$   & & $6.460$ & $-0.026~~(-4.0\text{\textperthousand})$ & & & $~~~\,\,\,0.131$ \\
with $\text{MA}$-$\text{Pb}$ & & $6.462$ & $-0.024~~(-3.8\text{\textperthousand})$ & & & $~~~\,\,\,0.122$ \\
with $\text{MA}$-$\text{I}$  & & $6.431$ & $-0.055~~(-8.4\text{\textperthousand})$ & & & $~~~\,\,\,0.273$ \\
with $\text{Pb}$-$\text{I}$  & & $6.476$ & $-0.010~~(-1.6\text{\textperthousand})$ & & & $~~~\,\,\,0.051$ \\
\textbf{Sum}                 & & $6.360$ & $-0.126(-19.4\text{\textperthousand})$  & & & \\ \hline
\multicolumn{7}{l}{$\text{CsPbI}_3^{}$} \\
PBE                          & &         &                                         & & $\alpha^{\text{PBE}}$ & $\beta^{\text{PBE}}$ \\
                             & & $6.398$ &                                         & & $3.1723$ & $-40.596$ \\
PBE+vdW                      & &         &                                         & & & $\beta^{\text{TS-vdW}}$ \\
with $\text{Cs}$-$\text{Cs}$ & & $6.359$ & $-0.039~~(-6.1\text{\textperthousand})$ & & & $~~~\,\,\,0.247$ \\
with $\text{Pb}$-$\text{Pb}$ & & $6.394$ & $-0.004~~(-0.6\text{\textperthousand})$ & & & $~~~\,\,\,0.026$ \\
with $\text{I}$-$\text{I}$   & & $6.372$ & $-0.026~~(-4.1\text{\textperthousand})$ & & & $~~~\,\,\,0.166$ \\
with $\text{Cs}$-$\text{Pb}$ & & $6.352$ & $-0.046~~(-7.2\text{\textperthousand})$ & & & $~~~\,\,\,0.291$ \\
with $\text{Cs}$-$\text{I}$  & & $6.291$ & $-0.107(-16.8\text{\textperthousand})$  & & & $~~~\,\,\,0.680$ \\
with $\text{Pb}$-$\text{I}$  & & $6.389$ & $-0.009~~(-1.4\text{\textperthousand})$ & & & $~~~\,\,\,0.059$ \\
\textbf{Sum}                 & & $6.167$ & $-0.232(-36.2\text{\textperthousand})$  & & & \\ \hline\hline
\end{tabular}
\end{center}
\end{table}

Compared with $\text{MAPbI}_3^{}$-\textbf{b}, $\text{CsPbI}_3^{}$ exhibits a narrower PBE total-energy curve (corresponding to a larger $\alpha^{\text{PBE}}$ value), as well as slightly larger gradients for $\Delta E_{\text{total}}^{\text{TS-vdW}}(\text{Pb-Pb})$, $\Delta E_{\text{total}}^{\text{TS-vdW}}(\text{I-I})$ and $\Delta E_{\text{total}}^{\text{TS-vdW}}(\text{Pb-I})$. The $\text{PbI}_3^-$ contribution to the lattice-constant reduction in these two systems is very similar. The major difference arises from the $\text{A}^+$ cation, as the gradients of all $\text{Cs}^+$-based $\Delta E_{\text{total}}^{\text{TS-vdW}}$ lines are much larger than their $\text{MA}^+$ counterparts. In particular, $\text{Cs}$-$\text{Cs}$ causes a lattice-constant reduction by $0.039~\text{\AA}$ ($6.1\text{\textperthousand}$), while the contribution of $\text{MA}$-$\text{MA}$, as mentioned earlier, is almost negligible. The largest contribution ($1.7\%$) comes from $\text{Cs}$-$\text{I}$. Consequently, vdW interactions in $\text{CsPbI}_3^{}$ result in the largest lattice-constant reduction in all investigated systems, as shown in Table~\ref{structuralparameters}.

In Table~\ref{vdWpairwise} (and Table~S3), we have switched on the TS-vdW interaction of each inter-particle pair separately, that is, vdW interactions of all other pairs are switched off. By summing up all contributions of the inter-particle pairs, we reach a ``total correction'' $\sum_{\text{Y-Z}}\Delta a^*_{\text{TS-vdW}}(\text{Y-Z})$, and the corresponding ``optimized'' lattice constant $a^*=a^*_{\text{PBE}}-\sum_{\text{Y-Z}}\Delta a^*_{\text{TS-vdW}}(\text{Y-Z})$. In general, these $a^*$ values are quite close to the lattice constants optimized using PBE+vdW, indicating that the lattice-constant corrections from the TS-vdW interaction of different inter-particle pairs are approximately additive \cite{Note1}.

These findings are instructive for the design of new hybrid perovskites toward the target unit-cell volumes, which are closely related to the materials' electronic properties (as demonstrated by our test calculations). Comparing Fig.~\ref{vdWcorrection}a with Fig.~\ref{vdWcorrection}c (or the first and the second panel of Table~S3), we find that the $E_{\text{total}}^{\text{PBE}}$ \textit{vs.} $a$ curve of $\text{MAPbI}_3^{}$-\textbf{a} is much narrower than that of $\text{MASnI}_3^{}$-\textbf{a}. As a result, vdW interactions with similar $\beta^{\text{TS-vdW}}$ parameters (such as $\text{I}$-$\text{I}$) have less impact on the lattice constant in $\text{MAPbI}_3^{}$-\textbf{a} than they do in $\text{MASnI}_3^{}$-\textbf{a}. Conversely, we have performed test calculations for $\text{CF}_3^{}\text{NH}_3^{}\text{PbI}_3^{}$-\textbf{a}. Compared with the isostructural $\text{CH}_3^{}\text{NH}_3^+$ cation, the trifluoride $\text{CF}_3^{}\text{NH}_3^+$ cation is subject to stronger vdW interactions due to the larger number of electrons. TS-vdW reduces the lattice constant of $\text{CF}_3^{}\text{NH}_3^{}\text{PbI}_3^{}$-\textbf{a} by $0.199~\text{\AA}$ (from $6.629$ to $6.430~\text{\AA}$). This effect is larger than in $\text{CH}_3^{}\text{NH}_3^{}\text{PbI}_3^{}$-\textbf{a} ($0.127~\text{\AA}$, see Table~\ref{latticeconstants}). To systematically exploit this effect for materials design, however, we would need to explore hybrid perovskites with many different compositions.

\section{Conclusions}\label{conclusions}

We have studied the atomic structure of a series of organometal-halide perovskites using DFT focussing in particular on the interaction between the organic cation and the inorganic matrix. We identify two stable configurations of the organic cation and analyse the associated deformation of the inorganic-framework in detail. The incorporation of vdW interactions into semi-local DFT calculations significantly corrects the calculated lattice constants and thus indirectly but strongly affects the atomic structure of hybrid perovskites. We further analyse the individual vdW contributions and identified the $\text{MA}$-$\text{I}$, $\text{I}$-$\text{I}$ and $\text{MA}$-$\text{Pb}$ interactions with the largest effect on the lattice constants. Our analysis of the vdW contributions provides insight into the design of new hybrid perovskites with favorable structural properties. This work serves as a foundation for future studies aiming at a supercell description of hybrid perovskites that reveals the tilting of the metal-halide octahedra and the alignment of organic cations in these systems as well as their mutual interplay.

\section*{Acknowledgment}

We thank H.~Levard, M.~Puska and K.~Laasonen as well as A.~Tkatchenko, V.~Blum and A.~Gulans for fruitful discussions. The generous allocation of computing resources by the CSC-IT Center for Science (via the Project No. ay6311) and the Aalto Science-IT project are gratefully acknowledged. This work was supported by the Academy of Finland through its Centres of Excellence Programme (2012-2014 and 2015-2017) under project numbers 251748 and 284621.

\section*{Associated content}

Supplemental Material: TS-vdW parameters for each atom; full coordinates of the investigated systems; vdW-caused lattice-constant corrections for $\text{MAPbI}_3^{}$-\textbf{a}, $\text{MASnI}_3^{}$-\textbf{a} and $\text{MAPbCl}_3^{}$-\textbf{a}. This material is available free of charge via the internet at http://journals.aps.org.

\bibliographystyle{apsrev}

\clearpage

\def\thepage{S\arabic{page}}
\setcounter{page}{0}
\begin{table*}[!ht]
\begin{center}
\textbf{Supplemental Material}
\end{center}
\end{table*}

\def\thetable{S\arabic{table}}
\setcounter{table}{0}

\begin{table*}[!ht]
\caption{TS-vdW parameters (density-scaled atomic polarizability $\alpha$, $C_6^{}$ coefficient and vdW radius $R_0^{}$) for each atom considered in this work. All data are in atomic units.}
\begin{center}
\begin{tabular}{lcccccc} \hline\hline
& $~$ & $\alpha$ & $~$ & $C_6^{}$ & $~$ & $R_0^{}$ \\ \hline
$\text{H}$  & & $~\,~\,4.50$ & & $~\,~\,~\,6.500$ & & $3.100$ \\
$\text{C}$  & & $~\,12.00$   & & $~\,~\,46.600$   & & $3.590$ \\
$\text{N}$  & & $~\,~\,7.40$ & & $~\,~\,24.200$   & & $3.340$ \\
$\text{Cl}$ & & $~\,15.00$   & & $~\,~\,94.600$   & & $3.710$ \\
$\text{Sn}$ & & $~\,55.95$   & & $~\,587.417$     & & $4.303$ \\
$\text{I}$  & & $~\,35.00$   & & $~\,385.000$     & & $4.170$ \\
$\text{Cs}$ & & $427.12$     & & $6582.080$       & & $3.780$ \\
$\text{Pb}$ & & $~\,61.80$   & & $~\,697.000$     & & $4.310$ \\ \hline\hline
\\
\\
\\
\\
\\
\\
\\
\\
\\
\\
\\
\\
\\
\\
\\
\\
\\
\\
\\
\\
\\
\\
\\
\\
\\
\\
\\
\\
\\
\\
\\
\\
\\
\\
\\
\\
\end{tabular}
\end{center}
\end{table*}

\begin{table*}[!ht]
\caption{Fractional coordinates of all inequivalent nuclei in optimized cubic primitive cells of all investigated hybrid perovskites: $\text{MAPbI}_3^{}$-\textbf{a}/\textbf{b}, $\text{MASnI}_3^{}$-\textbf{a}/\textbf{b}, and $\text{MAPbCl}_3^{}$-\textbf{a}/\textbf{b}. Both results calculated with PBE and PBE+vdW are listed.}
\begin{center}
\begin{tabular}{lcccccccccccccccc} \hline\hline
& $~$ & \multicolumn{3}{c}{PBE} & $~$ & \multicolumn{3}{c}{PBE+vdW} & $~$ & \multicolumn{3}{c}{PBE} & $~$ & \multicolumn{3}{c}{PBE+vdW} \\
& & $x$ & $y$ & $z$ & & $x$ & $y$ & $z$ & & $x$ & $y$ & $z$ & & $x$ & $y$ & $z$ \\ \hline
& & \multicolumn{7}{l}{$\text{MAPbI}_3^{}$-\textbf{a}} & & \multicolumn{7}{l}{$\text{MAPbI}_3^{}$-\textbf{b}} \\
$\text{C}$  & & $~~\,\,0.464$ & $~~\,\,0.464$ & $~~\,\,0.464$ & & $~~\,\,0.462$ & $~~\,\,0.462$ & $~~\,\,0.462$
            & & $~~\,\,0.446$ & $~~\,\,0.500$ & $~~\,\,0.498$ & & $~~\,\,0.440$ & $~~\,\,0.500$ & $~~\,\,0.468$ \\
$\text{H}$  & & $~~\,\,0.367$ & $~~\,\,0.563$ & $~~\,\,0.367$ & & $~~\,\,0.364$ & $~~\,\,0.563$ & $~~\,\,0.364$
            & & $~~\,\,0.376$ & $~~\,\,0.361$ & $~~\,\,0.565$ & & $~~\,\,0.357$ & $~~\,\,0.359$ & $~~\,\,0.521$ \\
$\text{H}$  & & $~~\,\,0.367$ & $~~\,\,0.367$ & $~~\,\,0.563$ & & $~~\,\,0.364$ & $~~\,\,0.364$ & $~~\,\,0.563$
            & & $~~\,\,0.428$ & $~~\,\,0.500$ & $~~\,\,0.331$ & & $~~\,\,0.454$ & $~~\,\,0.500$ & $~~\,\,0.297$ \\
$\text{H}$  & & $~~\,\,0.563$ & $~~\,\,0.367$ & $~~\,\,0.367$ & & $~~\,\,0.563$ & $~~\,\,0.364$ & $~~\,\,0.364$
            & & $~~\,\,0.376$ & $~~\,\,0.638$ & $~~\,\,0.565$ & & $~~\,\,0.357$ & $~~\,\,0.641$ & $~~\,\,0.521$ \\
$\text{N}$  & & $~~\,\,0.596$ & $~~\,\,0.596$ & $~~\,\,0.596$ & & $~~\,\,0.597$ & $~~\,\,0.597$ & $~~\,\,0.597$
            & & $~~\,\,0.670$ & $~~\,\,0.500$ & $~~\,\,0.549$ & & $~~\,\,0.656$ & $~~\,\,0.500$ & $~~\,\,0.561$ \\
$\text{H}$  & & $~~\,\,0.693$ & $~~\,\,0.512$ & $~~\,\,0.693$ & & $~~\,\,0.696$ & $~~\,\,0.512$ & $~~\,\,0.696$
            & & $~~\,\,0.744$ & $~~\,\,0.630$ & $~~\,\,0.491$ & & $~~\,\,0.741$ & $~~\,\,0.632$ & $~~\,\,0.516$ \\
$\text{H}$  & & $~~\,\,0.693$ & $~~\,\,0.693$ & $~~\,\,0.512$ & & $~~\,\,0.696$ & $~~\,\,0.696$ & $~~\,\,0.512$
            & & $~~\,\,0.698$ & $~~\,\,0.500$ & $~~\,\,0.707$ & & $~~\,\,0.656$ & $~~\,\,0.500$ & $~~\,\,0.724$ \\
$\text{H}$  & & $~~\,\,0.512$ & $~~\,\,0.693$ & $~~\,\,0.693$ & & $~~\,\,0.512$ & $~~\,\,0.696$ & $~~\,\,0.696$
            & & $~~\,\,0.744$ & $~~\,\,0.370$ & $~~\,\,0.491$ & & $~~\,\,0.741$ & $~~\,\,0.368$ & $~~\,\,0.516$ \\
$\text{Pb}$ & & $~~\,\,0.000$ & $~~\,\,0.000$ & $~~\,\,0.000$ & & $~~\,\,0.000$ & $~~\,\,0.000$ & $~~\,\,0.000$
            & & $~~\,\,0.000$ & $~~\,\,0.000$ & $~~\,\,0.000$ & & $~~\,\,0.000$ & $~~\,\,0.000$ & $~~\,\,0.000$ \\
$\text{I}$  & & $-0.016$      & $~~\,\,0.487$ & $-0.016$      & & $-0.006$      & $~~\,\,0.496$ & $-0.007$
            & & $-0.044$      & $~~\,\,0.500$ & $~~\,\,0.025$ & & $-0.024$      & $~~\,\,0.500$ & $~~\,\,0.028$ \\
$\text{I}$  & & $-0.016$      & $-0.016$      & $~~\,\,0.487$ & & $-0.007$      & $-0.006$      & $~~\,\,0.496$
            & & $-0.054$      & $~~\,\,0.000$ & $~~\,\,0.495$ & & $-0.047$      & $~~\,\,0.000$ & $~~\,\,0.498$ \\
$\text{I}$  & & $~~\,\,0.487$ & $-0.016$      & $-0.016$      & & $~~\,\,0.496$ & $-0.007$      & $-0.006$
            & & $~~\,\,0.472$ & $~~\,\,0.000$ & $-0.030$      & & $~~\,\,0.486$ & $~~\,\,0.000$ & $-0.031$ \\ \hline\hline
& & \multicolumn{7}{l}{$\text{MASnI}_3^{}$-\textbf{a}} & & \multicolumn{7}{l}{$\text{MASnI}_3^{}$-\textbf{b}} \\
$\text{C}$  & & $~~\,\,0.461$ & $~~\,\,0.461$ & $~~\,\,0.461$ & & $~~\,\,0.452$ & $~~\,\,0.452$ & $~~\,\,0.452$
            & & $~~\,\,0.441$ & $~~\,\,0.500$ & $~~\,\,0.481$ & & $~~\,\,0.430$ & $~~\,\,0.500$ & $~~\,\,0.464$ \\
$\text{H}$  & & $~~\,\,0.364$ & $~~\,\,0.560$ & $~~\,\,0.364$ & & $~~\,\,0.352$ & $~~\,\,0.555$ & $~~\,\,0.352$
            & & $~~\,\,0.369$ & $~~\,\,0.360$ & $~~\,\,0.546$ & & $~~\,\,0.346$ & $~~\,\,0.356$ & $~~\,\,0.517$ \\
$\text{H}$  & & $~~\,\,0.364$ & $~~\,\,0.364$ & $~~\,\,0.560$ & & $~~\,\,0.352$ & $~~\,\,0.352$ & $~~\,\,0.554$
            & & $~~\,\,0.429$ & $~~\,\,0.500$ & $~~\,\,0.311$ & & $~~\,\,0.445$ & $~~\,\,0.500$ & $~~\,\,0.290$ \\
$\text{H}$  & & $~~\,\,0.560$ & $~~\,\,0.364$ & $~~\,\,0.364$ & & $~~\,\,0.554$ & $~~\,\,0.352$ & $~~\,\,0.352$
            & & $~~\,\,0.369$ & $~~\,\,0.640$ & $~~\,\,0.546$ & & $~~\,\,0.346$ & $~~\,\,0.644$ & $~~\,\,0.517$ \\
$\text{N}$  & & $~~\,\,0.594$ & $~~\,\,0.594$ & $~~\,\,0.594$ & & $~~\,\,0.589$ & $~~\,\,0.589$ & $~~\,\,0.589$
            & & $~~\,\,0.667$ & $~~\,\,0.500$ & $~~\,\,0.539$ & & $~~\,\,0.649$ & $~~\,\,0.500$ & $~~\,\,0.558$ \\
$\text{H}$  & & $~~\,\,0.691$ & $~~\,\,0.509$ & $~~\,\,0.691$ & & $~~\,\,0.689$ & $~~\,\,0.502$ & $~~\,\,0.689$
            & & $~~\,\,0.744$ & $~~\,\,0.631$ & $~~\,\,0.482$ & & $~~\,\,0.736$ & $~~\,\,0.634$ & $~~\,\,0.513$ \\
$\text{H}$  & & $~~\,\,0.691$ & $~~\,\,0.691$ & $~~\,\,0.509$ & & $~~\,\,0.689$ & $~~\,\,0.689$ & $~~\,\,0.502$
            & & $~~\,\,0.691$ & $~~\,\,0.500$ & $~~\,\,0.700$ & & $~~\,\,0.648$ & $~~\,\,0.500$ & $~~\,\,0.724$ \\
$\text{H}$  & & $~~\,\,0.509$ & $~~\,\,0.691$ & $~~\,\,0.691$ & & $~~\,\,0.502$ & $~~\,\,0.689$ & $~~\,\,0.689$
            & & $~~\,\,0.744$ & $~~\,\,0.369$ & $~~\,\,0.482$ & & $~~\,\,0.736$ & $~~\,\,0.366$ & $~~\,\,0.513$ \\
$\text{Sn}$ & & $~~\,\,0.000$ & $~~\,\,0.000$ & $~~\,\,0.000$ & & $~~\,\,0.000$ & $~~\,\,0.000$ & $~~\,\,0.000$
            & & $~~\,\,0.000$ & $~~\,\,0.000$ & $~~\,\,0.000$ & & $~~\,\,0.000$ & $~~\,\,0.000$ & $~~\,\,0.000$ \\
$\text{I}$  & & $-0.019$      & $~~\,\,0.466$ & $-0.019$      & & $-0.009$      & $~~\,\,0.489$ & $-0.009$
            & & $-0.040$      & $~~\,\,0.500$ & $~~\,\,0.016$ & & $-0.022$      & $~~\,\,0.500$ & $~~\,\,0.022$ \\
$\text{I}$  & & $-0.019$      & $-0.019$      & $~~\,\,0.466$ & & $-0.009$      & $-0.009$      & $~~\,\,0.489$
            & & $-0.047$      & $~~\,\,0.000$ & $~~\,\,0.485$ & & $-0.043$      & $~~\,\,0.000$ & $~~\,\,0.498$ \\
$\text{I}$  & & $~~\,\,0.466$ & $-0.019$      & $-0.019$      & & $~~\,\,0.489$ & $-0.009$      & $-0.009$
            & & $~~\,\,0.458$ & $~~\,\,0.000$ & $-0.025$      & & $~~\,\,0.478$ & $~~\,\,0.000$ & $-0.023$ \\ \hline
& & \multicolumn{7}{l}{$\text{MAPbCl}_3^{}$-\textbf{a}} & & \multicolumn{7}{l}{$\text{MAPbCl}_3^{}$-\textbf{b}} \\
$\text{C}$  & & $~~\,\,0.452$ & $~~\,\,0.452$ & $~~\,\,0.452$ & & $~~\,\,0.443$ & $~~\,\,0.443$ & $~~\,\,0.443$
            & & $~~\,\,0.429$ & $~~\,\,0.500$ & $~~\,\,0.476$ & & $~~\,\,0.413$ & $~~\,\,0.500$ & $~~\,\,0.459$ \\
$\text{H}$  & & $~~\,\,0.344$ & $~~\,\,0.561$ & $~~\,\,0.344$ & & $~~\,\,0.333$ & $~~\,\,0.554$ & $~~\,\,0.333$
            & & $~~\,\,0.346$ & $~~\,\,0.345$ & $~~\,\,0.543$ & & $~~\,\,0.322$ & $~~\,\,0.343$ & $~~\,\,0.519$ \\
$\text{H}$  & & $~~\,\,0.344$ & $~~\,\,0.344$ & $~~\,\,0.561$ & & $~~\,\,0.333$ & $~~\,\,0.333$ & $~~\,\,0.554$
            & & $~~\,\,0.423$ & $~~\,\,0.500$ & $~~\,\,0.288$ & & $~~\,\,0.424$ & $~~\,\,0.500$ & $~~\,\,0.268$ \\
$\text{H}$  & & $~~\,\,0.561$ & $~~\,\,0.344$ & $~~\,\,0.344$ & & $~~\,\,0.554$ & $~~\,\,0.333$ & $~~\,\,0.333$
            & & $~~\,\,0.346$ & $~~\,\,0.655$ & $~~\,\,0.543$ & & $~~\,\,0.322$ & $~~\,\,0.657$ & $~~\,\,0.519$ \\
$\text{N}$  & & $~~\,\,0.598$ & $~~\,\,0.598$ & $~~\,\,0.598$ & & $~~\,\,0.592$ & $~~\,\,0.592$ & $~~\,\,0.592$
            & & $~~\,\,0.674$ & $~~\,\,0.500$ & $~~\,\,0.550$ & & $~~\,\,0.654$ & $~~\,\,0.500$ & $~~\,\,0.556$ \\
$\text{H}$  & & $~~\,\,0.706$ & $~~\,\,0.506$ & $~~\,\,0.706$ & & $~~\,\,0.701$ & $~~\,\,0.498$ & $~~\,\,0.701$
            & & $~~\,\,0.760$ & $~~\,\,0.645$ & $~~\,\,0.491$ & & $~~\,\,0.747$ & $~~\,\,0.647$ & $~~\,\,0.504$ \\
$\text{H}$  & & $~~\,\,0.706$ & $~~\,\,0.706$ & $~~\,\,0.506$ & & $~~\,\,0.701$ & $~~\,\,0.701$ & $~~\,\,0.498$
            & & $~~\,\,0.693$ & $~~\,\,0.500$ & $~~\,\,0.728$ & & $~~\,\,0.659$ & $~~\,\,0.500$ & $~~\,\,0.737$ \\
$\text{H}$  & & $~~\,\,0.506$ & $~~\,\,0.706$ & $~~\,\,0.706$ & & $~~\,\,0.498$ & $~~\,\,0.701$ & $~~\,\,0.701$
            & & $~~\,\,0.760$ & $~~\,\,0.355$ & $~~\,\,0.491$ & & $~~\,\,0.747$ & $~~\,\,0.353$ & $~~\,\,0.504$ \\
$\text{Pb}$ & & $~~\,\,0.000$ & $~~\,\,0.000$ & $~~\,\,0.000$ & & $~~\,\,0.000$ & $~~\,\,0.000$ & $~~\,\,0.000$
            & & $~~\,\,0.000$ & $~~\,\,0.000$ & $~~\,\,0.000$ & & $~~\,\,0.000$ & $~~\,\,0.000$ & $~~\,\,0.000$ \\
$\text{Cl}$ & & $-0.019$      & $~~\,\,0.484$ & $-0.019$      & & $-0.018$      & $~~\,\,0.492$ & $-0.018$
            & & $-0.046$      & $~~\,\,0.500$ & $~~\,\,0.026$ & & $-0.041$      & $~~\,\,0.500$ & $~~\,\,0.025$ \\
$\text{Cl}$ & & $-0.019$      & $-0.019$      & $~~\,\,0.484$ & & $-0.018$      & $-0.018$      & $~~\,\,0.492$
            & & $-0.051$      & $~~\,\,0.000$ & $~~\,\,0.492$ & & $-0.056$      & $~~\,\,0.000$ & $~~\,\,0.495$ \\
$\text{Cl}$ & & $~~\,\,0.484$ & $-0.019$      & $-0.019$      & & $~~\,\,0.492$ & $-0.018$      & $-0.018$
            & & $~~\,\,0.481$ & $~~\,\,0.000$ & $-0.037$      & & $~~\,\,0.491$ & $~~\,\,0.000$ & $-0.038$ \\ \hline\hline
\end{tabular}
\end{center}
\end{table*}

\clearpage

\begin{table*}[!ht]
\caption{Effects of the TS-vdW interactions for each particle pair for $\text{MAPbI}_3^{}$-\textbf{a}, $\text{MASnI}_3^{}$-\textbf{a} and $\text{MAPbCl}_3^{}$-\textbf{a}. For each system, we fit the second-order polynomial $E_{\text{total}}^{\text{PBE}}=\alpha^{\text{PBE}}a^2+\beta^{\text{PBE}}a+\gamma^{\text{PBE}}$ to the total energy calculated with PBE and list $\alpha^{\text{PBE}}$ and $\beta^{\text{PBE}}$. We also fit the linear function $\Delta E_{\text{total}}^{\text{TS-vdW}}=\beta^{\text{TS-vdW}}a+\gamma^{\text{TS-vdW}}$ to each vdW total-energy correction. Also listed are the optimized lattice constant $a^*$ calculated for each pair and its deviation $\Delta a^*$ to the PBE result. For the PBE+vdW results, all listed $\Delta a^*$-deviations are summed up to a value shown in the ``Sum'' row.}
\begin{center}
\begin{tabular}{lcccccc} \hline\hline
& $~$ & $a^*~[\text{\AA}]$ & $\Delta a^*~[\text{\AA}]$ & $~$ & \multicolumn{2}{c}{Fitting parameters} \\ \hline
\multicolumn{7}{l}{$\text{MAPbI}_3^{}$-\textbf{a}} \\
PBE                          & &         &                                         & & $\alpha^{\text{PBE}}$ & $\beta^{\text{PBE}}$ \\
                             & & $6.493$ &                                         & & $2.7247$ & $-35.382$ \\
PBE+vdW                      & &         &                                         & & & $\beta^{\text{TS-vdW}}$ \\
with $\text{MA}$-$\text{MA}$ & & $6.488$ & $-0.005~~(-0.8\text{\textperthousand})$ & & & $~~~\,\,\,0.027$ \\
with $\text{Pb}$-$\text{Pb}$ & & $6.489$ & $-0.004~~(-0.6\text{\textperthousand})$ & & & $~~~\,\,\,0.023$ \\
with $\text{I}$-$\text{I}$   & & $6.460$ & $-0.033~~(-5.1\text{\textperthousand})$ & & & $~~~\,\,\,0.181$ \\
with $\text{MA}$-$\text{Pb}$ & & $6.469$ & $-0.024~~(-3.6\text{\textperthousand})$ & & & $~~~\,\,\,0.128$ \\
with $\text{MA}$-$\text{I}$  & & $6.447$ & $-0.046~~(-7.1\text{\textperthousand})$ & & & $~~~\,\,\,0.250$ \\
with $\text{Pb}$-$\text{I}$  & & $6.484$ & $-0.009~~(-1.4\text{\textperthousand})$ & & & $~~~\,\,\,0.049$ \\
\textbf{Sum}                 & & $6.372$ & $-0.121(-18.6\text{\textperthousand})$  & & & \\ \hline
\multicolumn{7}{l}{$\text{MASnI}_3^{}$-\textbf{a}} \\
PBE                          & &         &                                         & & $\alpha^{\text{PBE}}$ & $\beta^{\text{PBE}}$ \\
                             & & $6.445$ &                                       & & $1.9335$ & $-24.921$ \\
PBE+vdW                      & &         &                                         & & & $\beta^{\text{TS-vdW}}$ \\
with $\text{MA}$-$\text{MA}$ & & $6.437$ & $-0.007~~(-1.2\text{\textperthousand})$ & & & $~~~\,\,\,0.029$ \\
with $\text{Sn}$-$\text{Sn}$ & & $6.440$ & $-0.005~~(-0.8\text{\textperthousand})$ & & & $~~~\,\,\,0.020$ \\
with $\text{I}$-$\text{I}$   & & $6.396$ & $-0.048~~(-7.5\text{\textperthousand})$ & & & $~~~\,\,\,0.187$ \\
with $\text{MA}$-$\text{Sn}$ & & $6.412$ & $-0.033~~(-5.1\text{\textperthousand})$ & & & $~~~\,\,\,0.127$ \\
with $\text{MA}$-$\text{I}$  & & $6.378$ & $-0.067(-10.4\text{\textperthousand})$  & & & $~~~\,\,\,0.260$ \\
with $\text{Sn}$-$\text{I}$  & & $6.434$ & $-0.011~~(-1.7\text{\textperthousand})$ & & & $~~~\,\,\,0.042$ \\
\textbf{Sum}                 & & $6.273$ & $-0.172(-26.7\text{\textperthousand})$  & & & \\ \hline
\multicolumn{7}{l}{$\text{MAPbCl}_3^{}$-\textbf{a}} \\
PBE$^*$                      & &         &                                         & & $\alpha^{\text{PBE}}$ & $\beta^{\text{PBE}}$ \\
                             & & $5.845$ &                                       & & $3.5074$ & $-40.998$ \\
PBE+vdW                      & &         &                                         & & & $\beta^{\text{TS-vdW}}$ \\
with $\text{MA}$-$\text{MA}$ & & $5.836$ & $-0.008~~(-1.4\text{\textperthousand})$ & & & $~~~\,\,\,0.057$ \\
with $\text{Pb}$-$\text{Pb}$ & & $5.839$ & $-0.006~~(-1.0\text{\textperthousand})$ & & & $~~~\,\,\,0.041$ \\
with $\text{Cl}$-$\text{Cl}$ & & $5.829$ & $-0.015~~(-2.6\text{\textperthousand})$ & & & $~~~\,\,\,0.106$ \\
with $\text{MA}$-$\text{Pb}$ & & $5.818$ & $-0.027~~(-4.6\text{\textperthousand})$ & & & $~~~\,\,\,0.187$ \\
with $\text{MA}$-$\text{Cl}$ & & $5.818$ & $-0.026~~(-4.5\text{\textperthousand})$ & & & $~~~\,\,\,0.183$ \\
with $\text{Pb}$-$\text{Cl}$ & & $5.837$ & $-0.007~~(-1.2\text{\textperthousand})$ & & & $~~~\,\,\,0.051$ \\
\textbf{Sum}                 & & $5.755$ & $-0.089(-15.3\text{\textperthousand})$  & & & \\ \hline\hline
\end{tabular}
\end{center}
\end{table*}

\end{document}